\documentclass[%
 reprint,
 superscriptaddress,
 amsmath,amssymb,
 aps,
 pra,
]{revtex4-1}

\usepackage{graphicx}
\usepackage{dcolumn}
\usepackage{bm}
\usepackage{natbib}
\usepackage{siunitx}
\usepackage[export]{adjustbox}
\usepackage{multirow}
\usepackage{accents}

\begin{document}


\title{Optimised fast gates for quantum computing with trapped ions}

\author{Evan P. G. Gale}
\affiliation{Department of Quantum Science, Research School of Physics, The Australian National University, Canberra, ACT 0200, Australia}
\author{Zain Mehdi}
\email{zain.mehdi@anu.edu.au}
\affiliation{Department of Quantum Science, Research School of Physics, The Australian National University, Canberra, ACT 0200, Australia}
\author{Lachlan M. Oberg}
\affiliation{Laser Physics Centre, Research School of Physics, The Australian National University, Canberra, ACT 0200, Australia}
\author{Alexander K. Ratcliffe}
\affiliation{Department of Quantum Science, Research School of Physics, The Australian National University, Canberra, ACT 0200, Australia}
\author{Simon A. Haine}
\affiliation{Department of Quantum Science, Research School of Physics, The Australian National University, Canberra, ACT 0200, Australia}
\author{Joseph J. Hope}
\affiliation{Department of Quantum Science, Research School of Physics, The Australian National University, Canberra, ACT 0200, Australia}

\date{\today}

\begin{abstract}
We present an efficient approach to optimising pulse sequences for implementing fast entangling two-qubit gates on trapped ion quantum information processors. We employ a two-phase procedure for optimising gate fidelity, which we demonstrate for multi-ion systems in linear Paul trap and microtrap architectures. The first phase involves a global optimisation over a computationally inexpensive cost function constructed under strong approximations of the gate dynamics. The second phase involves local optimisations that utilise a more precise ODE description of the gate dynamics, which captures the non-linearity of the Coulomb interaction and the effects of finite laser repetition rate. We propose two novel gate schemes that are compatible with this approach, and we demonstrate that they outperform existing schemes in terms of achievable gate speed and fidelity for feasible laser repetition rates. In optimising sub-microsecond gates in microtrap architectures, the proposed schemes achieve orders of magnitude higher fidelities than previous proposals. Finally, we investigate the impact of pulse imperfections on gate fidelity and evaluate error bounds for a range of gate speeds.
\end{abstract}


\maketitle


\section{\label{sec:introduction}Introduction}
A scalable quantum information processing (QIP) architecture would allow for the simulation of quantum systems that are too large to be tractable using classical computers \cite{Feynman1982, Spiller2005, Nielsen2010}. This would have significant benefits to the fields of quantum chemistry and fundamental physics \cite{Hempel2018, Lamata2014}, with these benefits flowing to the many fields that make use of these techniques and that study more complex compound systems. Numerous physical systems have been proposed for QIP architectures \cite{Ladd2010}, including photons \cite{Kok2007}, nuclear magnetic resonance \cite{Rong2017}, quantum dots \cite{Kloeffel2013}, nitrogen-vacancy centres in diamond~\cite{Childress2013}, superconducting circuits \cite{Makhlin2001}, and trapped ions \cite{Haffner2008}. \par

Trapped ion platforms are one of the most promising candidates for scalable QIP \cite{Steane1997, Wineland1998, Haffner2008, Nigg2014, Friis2018}. They have demonstrated excellent coherence properties \cite{Monroe2013}, near-perfect measurement readout \cite{Myerson2008}, and efficient implementation of entangling gates with low cross-talk \cite{Piltz2014}. Recent experiments have demonstrated single qubit rotations and multiple qubit entangling gates capable of conducting fault-tolerant quantum computation \cite{Harty2014, Gaebler2016, Ballance2016}, and a fully programmable five-qubit trapped ion quantum computer has been achieved \cite{Debnath2016}. While ion traps are capable of holding large numbers of qubits, their usefulness for quantum computation is fundamentally limited by the number of entangling operations that can be performed within the timescale for decoherence \cite{DiVincenzo1995a}. Therefore, the development of mechanisms for implementing high-speed and high-fidelity entangling gates is essential to realising large-scale computation on trapped ion platforms. \par

In this manuscript, we present methods for effective optimisation of pulse sequences that implement fast high-fidelity entangling gates on trapped ion quantum computers. In Section~\ref{sec:entangling_gates}, we describe the background of these gate mechanisms and existing approaches to their optimisation. In Section~\ref{sec:formalism_schemes}, we outline two novel gate schemes and propose methods for global optimisation of these pulse sequences. These schemes are applied to two-ion and multi-ion systems, which are compared to existing schemes in Section~\ref{sec:opt_results}. In Section~\ref{sec:local_opt}, we outline the secondary phase of optimisations, which provides a more complete description of the gate dynamics that includes the non-linearity of the Coloumb interaction and discretisation of the laser pulse timings. This two-phase procedure is summarised in Fig.~\ref{fig:Overview_Schematic}. Finally, in Section~\ref{sec:pulse_imperfections}, we investigate the impact of pulse imperfections on gate fidelity and evaluate the corresponding errors. \par
\begin{figure*}[t!]
    \centering
    \includegraphics[width=0.8\textwidth]{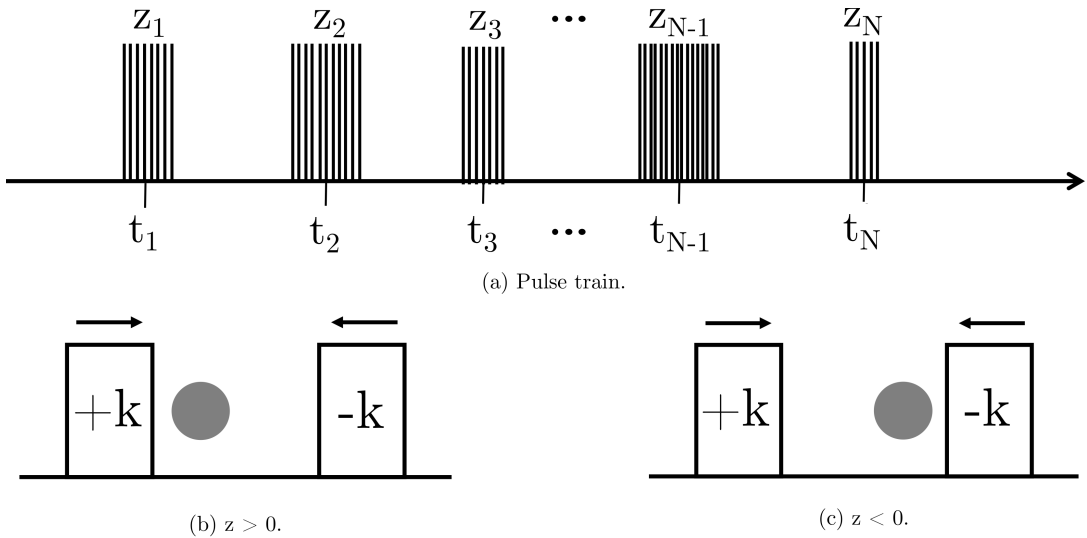}
    \caption{(a) Diagram of the pulse train for a general $N$-group fast gate pulse scheme where $z_j$ represents the number of pulse pairs in the pulse group arriving at the ion at time $t_j$. Each vertical line represents a single state-dependent kick. The sign of $z_j$ corresponds to which pulse of the counter-propagating pair arrives at the ion first, as shown in (b) and (c). This ultimately amounts to a sign change of the state-dependent momentum kick imparted on the ion. Adapted with permission from the supplementary material of Ratcliffe \textit{et al.}~\protect\cite{Ratcliffe2018}.}
    \label{fig:pulsetrain}
\end{figure*}
\section{\label{sec:entangling_gates}Background: Entangling gates in trapped ion QIP}
In ion trap experiments, ions are confined in a pseudo-harmonic potential generated by an oscillating electric field. The most common trapping architecture used is a radio-frequency Paul trap, in which multiple ions can be trapped in a linear array. Microtrap arrays are an alternative trapping architecture in which ions are confined in individual wells \cite{Ratcliffe2018, Cirac2000, Kumph2016}. The internal electronic states of the ions act as the qubits and are controlled through ion-light interactions with lasers. Entangling gates can be implemented by exploiting the connectivity of ions through the Coulomb interaction, via the motional states of the ions. \par

Almost all entangling gates that have been demonstrated to date involve addressing the motional sideband transitions, such as the Cirac-Zoller (CZ) \cite{Cirac1995, Jonathan2000, Schmidt-Kaler2003b, Schmidt-Kaler2003c} and M\o{}lmer-S\o{}rensen (MS) \cite{Molmer1999, Sorensen1999, Sorensen2000} gate schemes. However, sideband-resolving mechanisms require an adiabatic timescale to perform; the gate operation time must be significantly longer than the trapping period. Moreover, the presence of additional motional modes in systems of more than two ions results in sideband transitions that are harder to address, and thus lead to longer gate times as the number of ions is scaled \cite{Zhu2006}. Additionally, as more ions are added to a Paul trap, the trapping potential must be reduced along the longitudinal axis to prevent buckling of the linear ion chain \cite{Wineland1998}. This increases the trapping period, and in turn results in longer operation times for sideband-resolving entangling gates. \par

\subsection{\label{sec:fast_gates}Fast (non-adiabatic) entangling gates}
There have been a variety of proposals on alternate entangling gate mechanisms that are not subject to the adiabatic timescales that limit sideband-resolving gates. Originally proposed by Garc\'{i}a-Ripoll, Zoller and Cirac (GZC) \cite{Garcia-Ripoll2003, Garcia-Ripoll2005}, these mechanisms excite multiple motional modes during gate operation, as opposed to targeting individual sideband transitions. As a result, the gates can be implemented with speeds faster than the trapping period, for which they are often dubbed ``fast gates''. This proposal has recently been experimentally demonstrated by Sch\"{a}fer \textit{et al}.~\cite{Schafer2018}, who used an amplitude-shaped pulse to drive state-dependent displacements of both the common-motional and breathing modes of a two-ion system \cite{Steane2014}. They were able to demonstrate a high-fidelity ($99.8\%$) gate in \SI{1.6}{\micro\second}, and a sub-trap period \SI{480}{\nano\second} gate albeit with lower fidelity (${\sim}60\%$). \par

There have also been proposals for implementing fast gates using ultra-fast resonant laser pulses \cite{Garcia-Ripoll2003, Duan2004a, Bentley2013, Bentley2015b, Bentley2016, Ratcliffe2018, Hussain2016, Heinrich2019}, which have not yet been experimentally demonstrated and are the main focus of this manuscript. In these fast gate schemes, counter-propagating pairs of ultra-fast laser pulses impart state-dependent momentum kicks on pairs of ions. We assume that each counter-propagating pair is split from the same pulse with a very short delay engineered between the forward and counter-propagating pulses, although this is not strictly required. Sequences of these state-dependent kicks, interspersed with free evolution (see Fig.~\ref{fig:pulsetrain}), orchestrate state-dependent trajectories of the ions through phase space. The difference in the areas enclosed in the phase-space trajectories of the $\{|00\rangle, \, |11\rangle\}$ states and the $\{|10\rangle, \, |01\rangle\}$ states leads to acquisition of a relative phase. If this relative phase is $\frac{\pi}{2}$, and the motional states are disentangled by the end of the operation, the controlled-phase (CPH) gate is implemented. This gate has the unitary
\begin{align}
\label{eq:CP_unitary}
    \hat{U}_\text{CPH} = e^{i \frac{\pi}{4} \hat{Z}_1 \otimes \hat{Z}_2} \, ,
\end{align}
where $\hat{Z}_k$ is the Pauli-$Z$ operator acting on the $k$-th ion. This is a maximally entangling two-qubit gate and is equivalent to a CNOT operation, up to local rotations. To realise this ideal unitary operation on two qubits (labelled $A$ and $B$) with a series of state-dependent kicks from $N$ pulse groups, the following conditions must be satisfied:
\begin{gather}
\label{eq:phase_condition}
   \sum_p 8\eta^2\frac{\omega_t}{\omega_p} b_p^A b_p^B\sum_{i\neq j} z_i z_j \sin(\omega_p |t_i-t_j|) = \frac{\pi}{4} \, , \\
\label{eq:motional_restoration}
   2 \eta \sqrt{\frac{\omega_t}{\omega_p}} \sum_{k = 1}^{N} z_k e^{-i \omega_p t_k} = 0 \, ,
\end{gather}
where $z_k$ is the number of pulse pairs in the $k$-th pulse group which arrives at the ion at some time $t_k$, and $\omega_p$ is the angular frequency of the $p$-th motional mode \cite{Bentley2015b}. The element $b_p^i$ corresponds to the coupling of the position of the $i$-th ion to the $p$-th motional mode, i.e.
\begin{align} \label{eq:excursion}
    \hat{x}^i = \sum_p b_p^i \hat{Q}_p \, ,
\end{align}
where $\hat{Q}_p$ is the operator for the $p$-th motional mode. Eq.~\eqref{eq:excursion} is a normal mode expansion and describes small excursions of the ions from their equilibrium positions~\cite{Wineland1998}. We have defined the Lamb-Dicke parameter to be independent of the motional mode,
\begin{align}
    \eta \equiv \sqrt{\frac{\hbar}{2 M \omega_t}} \, ,
\end{align}
where $M$ is the mass of the ion species and $\omega_t$ is the angular trap frequency associated with the axis along which the gate is performed. Eq.~\eqref{eq:phase_condition} expresses the desired phase acquisition, while Eq.~\eqref{eq:motional_restoration} describes the disentangling of each motional mode from the internal states of the ions. Since all motional modes must be disentangled by the end of the operation, Eq.~\eqref{eq:motional_restoration} forms a set of $N_P$ conditions, where $N_P$ is the number of excited motional modes. Each of these conditions need to be addressed in order to realise a fast gate, which can be achieved by approaching the pulse sequence design as an optimisation problem~\cite{Glaser2015}. \par

Designing a particular pulse sequence to implement an entangling gate using this approach involves two sets of free parameters: the number of pulse pairs in each pulse group $z_k$, and the timings $t_k$ describing the arrival of the $k$-th pulse group at the ions. For a gate composed of $N$ pulse groups, this constitutes a $2N$-dimensional parameter space
\begin{align}
\begin{aligned}
    \vec{z} &= \{ z_1, \, z_2, \, z_3,\dots,z_N \} \, , \\
    \vec{t} &= \{ t_1, \, t_2, \, t_3,\dots, t_N \} \, .
\end{aligned}
\end{align}
The elements of $\vec{z}$ are integers and are able to take negative values. The sign of each $z_k$ corresponds to the direction of the state-dependent kick, i.e. the choice of which pulse arrives first in the counter-propagating pair, as shown in Fig.~\ref{fig:pulsetrain}. In general, optimisation over this unconstrained parameter space is intractable. For effective gate optimisation, constraints must be placed on these parameters to reduce the dimensionality of the search space. \par
\begin{figure*}
    \centering
    \includegraphics[width=\textwidth]{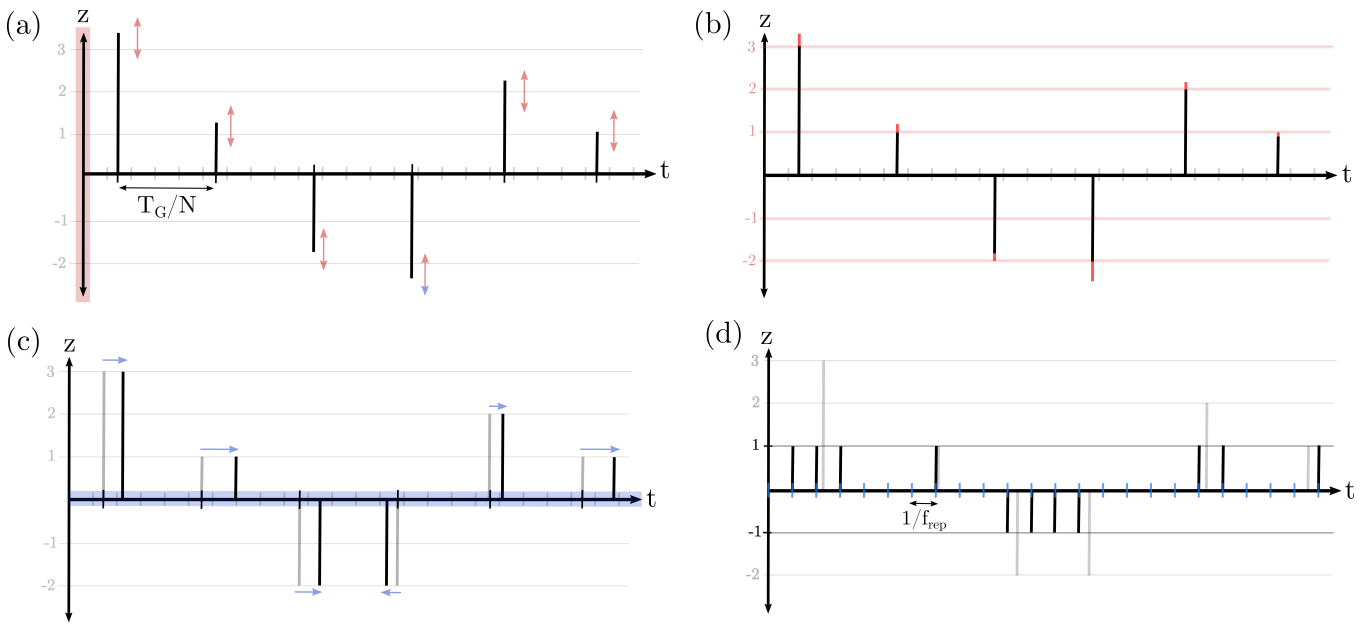}
    \caption{(Colour online). Schematic overview of the proposed two-phase optimisation procedure, for a simple pulse sequence with $N = 6$ pulse groups. (a) In the first phase of the optimisation (red), a global optimisation over the number of pulse pairs in each group is implemented using an approximated form of the infidelity, Eq.~\eqref{eq:truncatedinf}, as the cost function. Here, $T_G$ is the total gate time and $z_k$ is the number of pulse pairs in the $k$-th pulse group; a negative value corresponds to a change of sign in the pulse wavenumber. In this stage it is assumed that the timing between pairs of pulses in each group is negligible (i.e. infinite laser repetition rate). These elements are allowed to take non-integer values during global optimisation, after which (b) they are rounded to integer values. (c) In the second phase of the optimisation procedure (blue), the timings of the pulse groups are locally optimised using an ODE description of the motion of the ions, which includes the non-linearity of the Coulomb interaction. (d) The pulse groups are broken up into their constituent pairs, the timings of which are snapped to a grid defined by the laser repetition rate $1/f_\text{rep}$.}
    \label{fig:Overview_Schematic}
\end{figure*}
\subsection{\label{sec:constr_optimisations}Constraining fast gate optimisations}
Previous studies of these fast gate mechanisms have shown that arbitrarily fast two-qubit gate speeds can be achieved with high fidelity, given arbitrarily high laser repetition rates \cite{Duan2004a,Bentley2015b, Taylor2017}. Currently lasers can reach repetition rates on the order of \SI{300}{\mega\hertz} \cite{Hussain2016}, although \SI{5}{\giga\hertz} repetition rates have also recently been demonstrated \cite{Heinrich2019}. For achieving gate speeds faster than the trap period, the  difficulty is choosing the sequence of state-dependent kicks such that Eqs.~\eqref{eq:phase_condition} and \eqref{eq:motional_restoration} are satisfied, with pulse sequences that are compatible with available laser repetition rates. \par

For a given set of constraints (e.g. gate speed, total number of pulses) there is no known analytic solution for the optimal pulse sequence that best satisfies these equations. Instead, pulse sequences are identified through numerical optimisation, which can be done by minimising a cost function. We choose this cost function to be the infidelity of the gate, $1 - F$, where $F$ is the state-averaged fidelity measure \cite{Pedersen2008}. In previous work~\cite{Ratcliffe2018}, we have presented a truncated form of the infidelity, assuming the motional modes initially start in a thermal product state \cite{Taylor2017},
\begin{align}
    \label{eq:truncatedinf}
    1 - F \simeq \frac{2}{3} \Delta \phi^2 + \frac{4}{3}\sum_p \left( \frac{1}{2} + \Bar{n}_p \right) \big{(} (b_p^A)^2 + (b_p^B)^2 \big{)} \Delta P_p^2 \, ,
\end{align}
where $\Bar{n}_p$ is the average occupation number of the $p$-th motional mode. Here, $\Delta\phi$ is the phase-mismatch and $\Delta P_p$ is the unrestored motion of the $p$-th motional mode:
\begin{align}
    \label{eq:dphi expression}
    \Delta \phi &= \left| 8\eta^2 \sum_p \frac{\omega_t}{\omega_p} b_p^A b_p^B \sum_{i\neq j} z_i z_j \sin(\omega_p \left| t_i - t_j \right|) \right| - \frac{\pi}{4} \, , \\
    \label{eq:dPp expression}
    \Delta P_p &= 2\eta \sqrt{\frac{\omega_t}{\omega_p}} \left| \sum_{k = 1}^{N} z_k e^{-i \omega_p t_k} \right| \, .
\end{align}
Eq.~\eqref{eq:truncatedinf} provides a good approximation of the true infidelity and converges rapidly for high-fidelity gates. The expression is computationally cheap to evaluate and is therefore a suitable cost function for numerical optimisation procedures. In the case of perfect phase acquisition ($\Delta\phi = 0$) and perfect motional restoration ($\Delta P_p = 0, \; \forall p$), the ideal gate is implemented ($F = 1$). However, Eqs.~\eqref{eq:dphi expression} and \eqref{eq:dPp expression} assume an infinite laser repetition rate and a Coulomb interaction truncated to second order. These are strong assumptions that allow Eq.~\eqref{eq:truncatedinf} to be computed inexpensively, but will damage gate fidelity if not corrected for. In Section~\ref{sec:local_opt}, we describe a method that corrects for these assumptions as part of our proposed optimisation procedure. \par

The GZC scheme \cite{Garcia-Ripoll2003, Garcia-Ripoll2005}, as well as the fast robust anti-symmetric gate (FRAG) scheme proposed by Bentley \textit{et al.}~\cite{Bentley2013, Bentley2015b, Bentley2016}, constrained the parameter space for a gate with $N = 6$ pulse groups, by imposing fixed, anti-symmetric ratios for $z_k$. Both the GZC and FRAG schemes can be written in the form 
\begin{align}
\begin{aligned}
    \vec{z} &= n\{ -a, \, -b, \, -c, \, c, \, b, \, a \} \, , \\
    \vec{t} &= \{ -\tau_1, \, -\tau_2, \, -\tau_3, \, \tau_3, \, \tau_2, \, \tau_1 \} \, ,
\end{aligned}
\end{align}
where $(a,b,c)=(2,-3,\,2)$ for the GZC scheme and $(a,b,c)=(1,-2,\,2)$ for the FRAG scheme. These constraints reduce a na\"{i}ve optimisation over a twelve-dimensional space to just four dimensions $(n, \, \tau_1, \, \tau_2, \, \tau_3)$, which is far less expensive to search over. In the original FRAG scheme, a strict ordering of the pulse groups was imposed, i.e. $\tau_1>\tau_2>\tau_3$. This restriction was lifted in later analyses \cite{Taylor2017, Ratcliffe2018}, which allowed high-fidelity solutions to be found for a wider range of gate times and experimental parameters. \par

As both the GZC and FRAG schemes have three free parameters in time, gate solutions can be found that exactly satisfy the phase condition~\eqref{eq:phase_condition}, and the restoration of the two motional modes~\eqref{eq:motional_restoration}. However, when there are more than two relevant motional modes (e.g. in systems of more than two ions), the GZC and FRAG schemes are no longer capable of finding solutions that exactly satisfy Eqs.~\eqref{eq:phase_condition} and \eqref{eq:motional_restoration}. As a result, these schemes are not well suited for gate optimisations of multi-ion systems. \par

Furthermore, the infidelity expression is sinusoidal in the space of pulse group timings, i.e. $1 - F \sim \sin^2{(\omega_p t)}$, which is computationally expensive and thus is ill-suited for use as a cost function for global optimisation. An additional consequence of searching over pulse timings is that there is no intrinsic cost of pulse groups being arbitrarily close together in time, which may result in pulse sequences that require restrictively large repetition rates to implement. \par
 
An alternative gate scheme proposed by Duan \cite{Duan2004a} separates each pulse group out into individual pulse pairs, i.e. $|z_k| = 1 \; \forall k$, with the periods of free evolution given only by the finite repetition period $t_{rep} = 1/f_{rep}$ of the laser. The signs of the $z_k$ elements are chosen in clusters, such that closed loops are created by phase-space trajectories. In the simplest form of the Duan scheme, a single loop through phase space is orchestrated, resembling a triangular shape \cite{Bentley2015b}. Formally, the optimisation of the Duan scheme only involves a single free variable in time (determined by the total number of pulses used), which is typically chosen to satisfy the phase condition~\eqref{eq:phase_condition}. In order to achieve high-fidelity motional restoration, multiple loops through phase space are required, which in turn requires larger numbers of pulses and leads to longer gate times \cite{Duan2004a}. In the limit of infinite repetition rate, the Duan scheme is optimal as it utilises all laser pulses. However, the GZC and FRAG schemes generally result in faster and higher-fidelity gates for experimentally feasible laser repetition rates \cite{Bentley2015b}. \par
\begin{figure*}
    \centering
    \includegraphics[width=\textwidth]{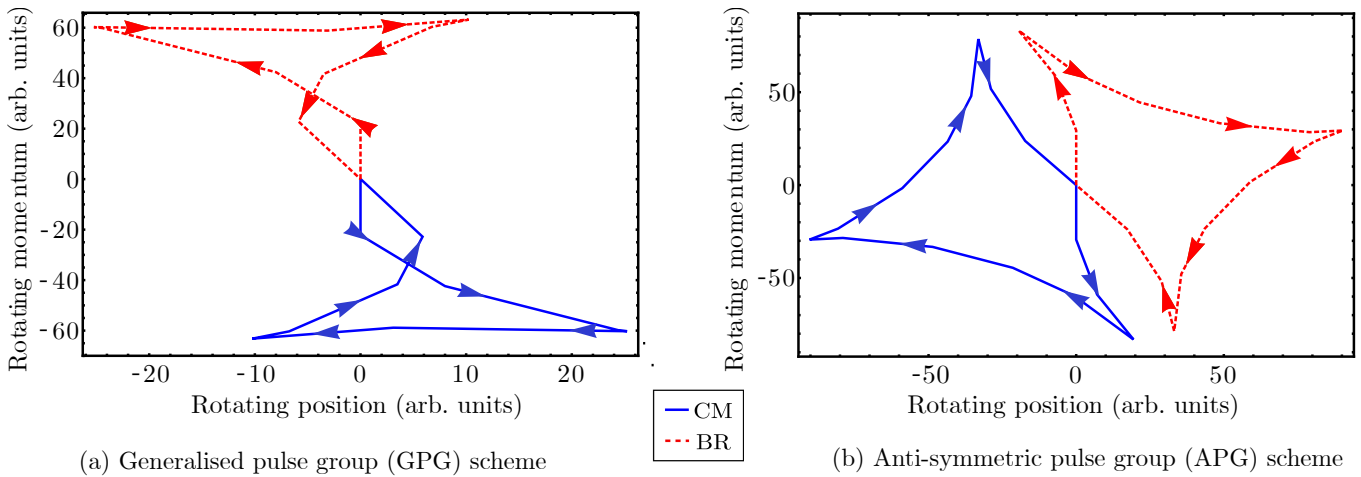}
    \caption{(Colour online). Trajectories through rotating phase space of the common-motional (blue, solid) and breathing (red, dotted) modes of a fast gate, exciting the motional modes along the longitudinal axis of a microtrap array. This is shown for a gate optimised with the (a) generalised pulse group (GPG) scheme, and (b) anti-symmetric pulse group (APG) scheme. In each case, both modes are clearly restored to the phase-space origin by the end of the gate operation.}
    \label{fig:phasespace}
\end{figure*}
In this manuscript, we present an efficient two-phase optimisation procedure for designing fast two-qubit entangling gates. In the first phase, global optimisation is implemented over elements of $\vec{z}$, with constraints placed on the pulse group timings $t_k$ using the linearised cost function~\eqref{eq:truncatedinf}. This is followed by a second phase of local optimisations on the pulse timings using an ordinary differential equation (ODE) description of the gate dynamics, which accounts for the non-linear effects of the Coulomb interaction and the finite repetition rate of the laser. An overview of our approach is presented in Fig.~\ref{fig:Overview_Schematic}. \par

\section{\label{sec:formalism_schemes}Global optimisation methods with arbitrary pulse groups}
We propose an alternative method for global optimisation of the cost function~\eqref{eq:truncatedinf}, where constraints are placed on the elements of $\vec{t}$ and the optimisation is performed over the elements $z_k$. The cost function is quartic in this space, i.e. $1 - F \sim \mathcal{O} \, \left( z_k^4 \right)$, and is thus computationally cheaper than the sinusoidal infidelity function involved in the FRAG and GZC schemes. Formally, this is an integer-programming problem as the parameters $z_k$ are integers corresponding to the discrete number of pulse pairs in each group. However, it is sufficient to treat these elements as continuous variables for the first phase of global optimisation, and round to the nearest integer values (see Fig.~\ref{fig:Overview_Schematic}). Further details of our numerical approach to global optimisation are provided in Appendix~\ref{app:global_opt}. \par

Specifically, we propose a scheme in which we initially assume that the pulse groups arrive at the ions at regularly spaced time intervals. For $N$ pulse groups, this scheme is written as
\begin{align}
\label{eq:GPG(N)_zandt}
\begin{aligned}
    \vec{z}_N &= \left\{ z_1, \, z_2, \, z_3, \, \dots, \, z_{N - 1}, \, z_N \right\} \, , \\
    \vec{t}_N &= \frac{T_G}{N} \left\{ 1, \, 2, \, \dots, \, N - 1,\, N \right\} \, ,
\end{aligned}
\end{align}
where $T_G$ is the gate operation time. We call this the generalised pulse group scheme, denoted as GPG$(N)$ for $N$ pulse groups. The regular spacing of pulse group timings in the GPG scheme ensures that resulting gate solutions are robust to finite laser repetition rates, as it minimises chances of pulse groups overlapping. It is unlikely that optimal fast gate schemes always conform to regularly spaced pulse groups, but this can be overcome with further local optimisations on the pulse group timings. These local optimisations are performed with an ODE description of the classical motion of the ions, which has the benefit of being able to include non-linearities of the Coulomb interaction and effects of finite repetition rate. We investigate this further in Section~\ref{sec:local_opt}. \par

Where the parameter space has additional complexity, it is convenient to impose further restrictions for computational ease. For example, in a system of three or more ions, or where the number of pulse groups is larger than $N \sim 14$, re-imposing anti-symmetric constraints (similar to GZC and FRAG) can be particularly useful. We propose a second set of constraints for global optimisation, which we call the anti-symmetric pulse group (APG) scheme; it has the form
\begin{align}
\label{eq:APG(N)_zandt}
\begin{aligned}
    \vec{z}_N &= \left\{ -z_{N/2}, \, \dots, \, -z_2, \, -z_1, \, z_1, \, z_2, \, \dots, \, z_{N/2} \right\} \, , \\
    \vec{t}_N &= \frac{T_G}{N} \left\{ -\frac{N}{2}, \, \dots, \, -2, \, -1, \, 1, \, 2, \, \dots, \, \frac{N}{2} \right\} \, .
\end{aligned}
\end{align}
These constraints reduce not only the dimensionality of the search space, but also the complexity of the cost function since momentum restoration of each motional mode is guaranteed. When anti-symmetric constraints are imposed, the general expression for motional restoration is
\begin{align}
    \label{eq:motres_antisym}
    \Delta P_p = 2\eta \sqrt{\frac{\omega_t}{\omega_p}} \sum_{k = 1}^{N} z_k \sin(\omega_p t_k) \, ,
\end{align}
\begin{figure*}
    \centering
    \includegraphics[width=\linewidth,valign=c]{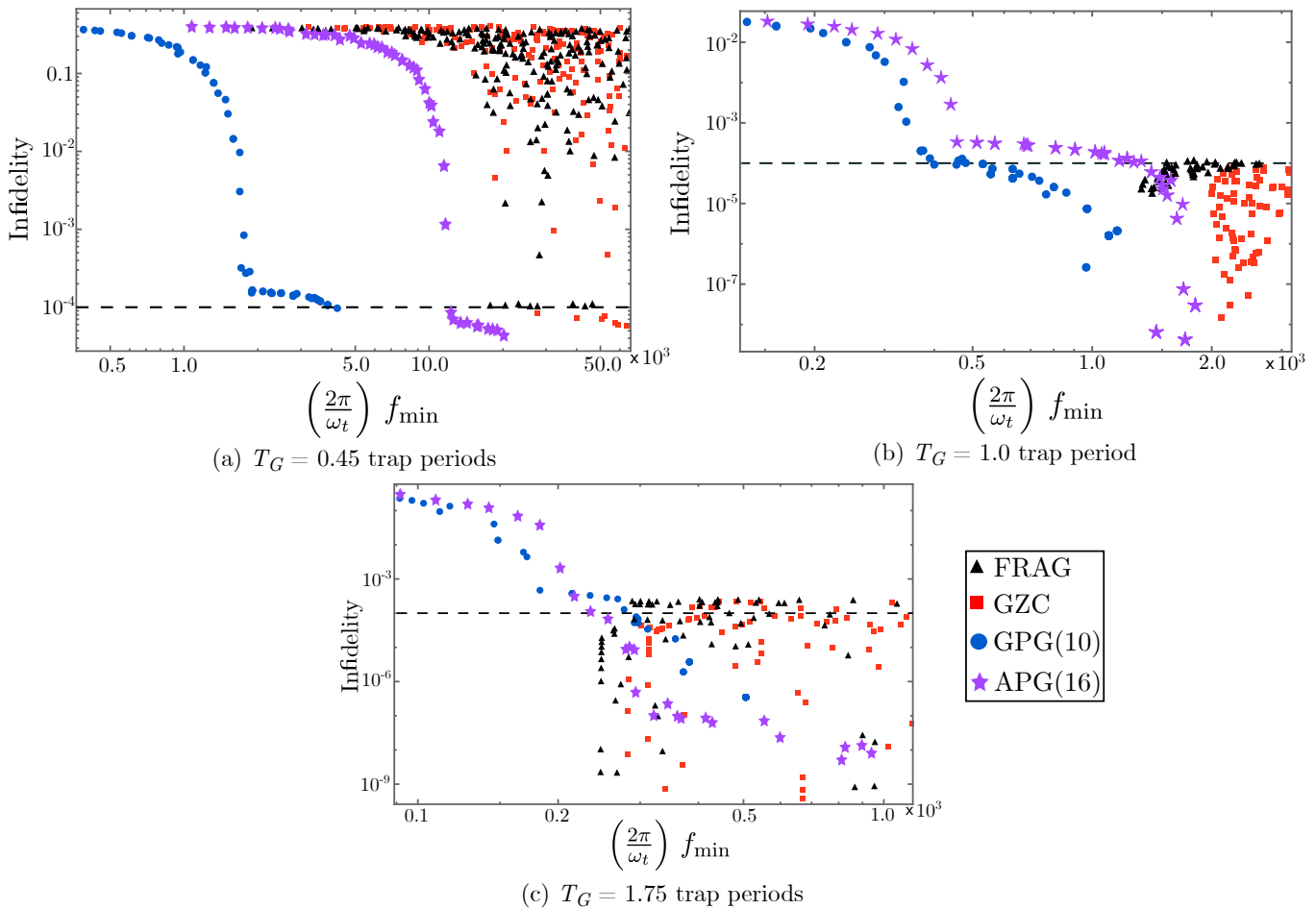} 
    \caption{(Colour online). Comparison of the FRAG, GZC, GPG$(10)$ and APG$(16)$ schemes for optimised fast gates between two ions in a linear microtrap chain for different gate times. We report the infidelities, calculated by the truncated cost function~\eqref{eq:truncatedinf}, with respect to the minimum repetition rate $f_\text{min}$ required to resolve the pulse groups, i.e. such that the pulse pairs do not overlap. The black dashed line marks an indicative infidelity threshold of $10^{-4}$.}
    \label{fig:MicrotrapOpt}
\end{figure*}
which is computationally cheaper than the full expression given in Eq.~\eqref{eq:motional_restoration}. Phase-space trajectories for exemplary gates optimised under the GPG and APG schemes are visualised in Fig.~\ref{fig:phasespace}. Formally, as the APG scheme is a restriction of the more general GPG scheme, it is unlikely to find higher fidelity solutions than the GPG scheme given sufficient computational resources. However, given the simplifications of the anti-symmetric constraints, the APG scheme is well suited to optimisations where the cost function is sufficiently complex that it cannot be effectively sampled in the allowed parameter space. We therefore use the APG scheme to optimise gates in ion chains with more than two ions, which we consider in Section~\ref{sec:LP_scaling}. \par

During the optimisation procedure, we assume a noiseless ion trap, and neglect motional heating and dephasing as potential sources of error. These mechanisms are typically much slower than typical fast gate dynamics which occur on the timescale set by the trapping period (${\sim}$\SI{1}{\micro\second}), and are highly unlikely to affect the dynamics of a single gate operation \cite{Taylor2017}. The optimisation method does not take into account imperfections in the laser control, which are of more concern to fast gate dynamics. Errors in pulse area arising from laser intensity fluctuations are particularly damaging to fast gate implementations, which we investigate in depth in Section~\ref{sec:pulse_imperfections}. \par
\begin{figure*}
    \centering
    \includegraphics[width=\linewidth]{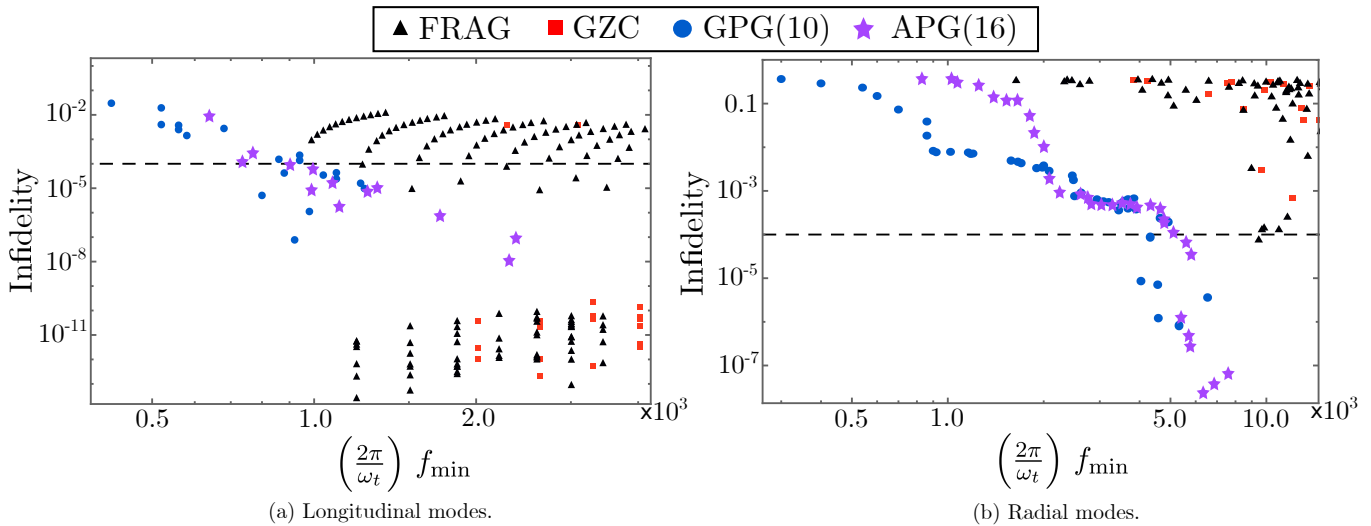}
    \caption{(Colour online). Comparison of the FRAG, GZC, GPG$(10)$ and APG$(16)$ schemes for a gate time of $T_G = 0.25$ trap periods. Optimisations are presented for a two-ion system in a common Paul trap, with state-dependent kicks performed on the (a) longitudinal and (b) radial trap modes. We report the infidelities with respect to the minimum resolving repetition rate, $f_\text{min}$. The black dashed line marks an indicative infidelity threshold of $10^{-4}$.}
    \label{fig:LP2ion}
\end{figure*}
\section{\label{sec:opt_results}Optimisation Results}
In this section, we present results of the global optimisation of fast gates using the GPG and APG schemes, which we compare to results of optimisations under the GZC and FRAG schemes. The speed of fast gate operations is limited by the available laser repetition rate, due to the scaling relation $T_G \propto f_{rep}^{-2/5}$ \cite{Bentley2013}, so we report the state-averaged infidelities against the minimum laser repetition rate $f_{min}$. This is the minimum repetition rate required to ensure that individual pulse groups in a given pulse scheme do not overlap in time, which is required for fast gates to be robust to the effects of finite repetition rates \cite{Bentley2015b, Taylor2017, Ratcliffe2018}. We investigate the explicit inclusion of finite repetition rate in the following section. \par

For all calculations presented in this manuscript, we consider $^{40}\text{Ca}^{+}$ as the candidate ion species, with corresponding parameters detailed in Appendix~\ref{app:param_choices}. We focus on gate speeds of the same order as, or faster than, the trapping period; in general, this requires repetition rates on the order of the state-of-the-art (\SI{5}{\giga\hertz} \cite{Heinrich2019}) to resolve. The results of this section can be extrapolated to slower gate speeds, which have lower repetition rate requirements via the aforementioned relation $T_G \propto f_{rep}^{-2/5}$. \par

\subsection{\label{sec:twoion_results} Two-ion systems}
For a simple two-ion system, fast gate dynamics can be entirely captured by the normalised difference between the two motional modes of the system, as given by the dimensionless parameter
\begin{align} \label{eq:mode_differences}
    \chi \equiv \frac{\omega_b - \omega_c}{\omega_t} \, ,
\end{align}
where $\omega_c = \omega_t$ and $\omega_b$ are the frequencies of the common-motional and breathing modes, respectively. The parameter $\chi$ can be calculated directly from fundamental trap parameters, as described in the supplementary material of Ratcliffe \textit{et al.}~\cite{Ratcliffe2018}. We will first consider fast gates between two ions in a linear microtrap design, with counter-propagating pulses parallel to the ion chain. This geometry is characterised by $\chi = \SI{1.8e-4}{}$, roughly corresponding to \SI{90}{\micro\meter} separation between the minima of microtraps with $\omega_t = 2\pi \times \SI{1.2}{\mega\hertz}$. In Fig.~\ref{fig:MicrotrapOpt}, we present global optimisations of fast gates in this microtrap architecture for gate times of $0.45$, $1.0$ and $1.75$ trap periods. \par

For sub-microsecond gates in a linear microtrap array, the FRAG and GZC schemes are only able to find low infidelity solutions with restrictively high repetition rates. Therefore, Fig.~\ref{fig:MicrotrapOpt}(a) provides a means to compare the GPG and APG schemes against the worst of FRAG and GZC. Fig.~\ref{fig:MicrotrapOpt} clearly shows that the GPG scheme in particular is able to find gate solutions with infidelities around $10^{-4}$ that can be resolved with ${\sim}\SI{2}{\giga\hertz}$ repetition rates (assuming $\frac{\omega_t}{2\pi} \sim \SI{1}{\mega\hertz}$), which falls well within experimental feasibility. \par

Another useful comparison is at $1$ trap period, shown in Fig.~\ref{fig:MicrotrapOpt}(b). FRAG and GZC optimisations are able to find solutions with high fidelity, but only with repetition rates larger than ${\sim}1500\;\frac{2\pi}{\omega_t}$. Both the GPG and APG schemes are able to reach very high fidelities for much lower repetition rates of $500\text{ -- }1000\;\frac{2\pi}{\omega_t}$. Fig.~\ref{fig:MicrotrapOpt}(c) shows that, at a gate time of $1.75$ trap periods, FRAG and GZC are able to find extremely high-fidelity gate solutions that are comparable to those found by APG and GPG optimisations, requiring repetition rates of only ${\sim}\SIrange[range-units=single, range-phrase=\text{ -- }]{100}{500}{\mega\hertz}$ to resolve. However, only the APG and GPG schemes are able to find solutions that are resolvable with repetition rates lower than $200\;\frac{2\pi}{\omega_t}$. \par

Much faster gate times are achievable for gates between ions in a common linear Paul trap, given the faster phase acquisition due to the significantly stronger Coulomb interaction at small inter-ion separation \cite{Bentley2015b}. The state-dependent kicks given to the ions during gate operation can be performed parallel or perpendicular to the ion chain, depending on the orientation of the laser pulses, which leads to excitation of the longitudinal and radial motional modes, respectively. \par

In Fig.~\ref{fig:LP2ion}, we present optimisations for two ions in a linear Paul trap for a gate time of $0.25$ trap periods. This is presented for gates performed both on the longitudinal modes~($\chi = \sqrt{3} - 1$) and on the radial modes~($\chi = -\SI{1.4e-2}{}$), in Fig.~\ref{fig:LP2ion}. In both cases, the APG and GPG schemes are shown to find high-fidelity solutions for lower laser repetition rates, which is particularly evident for gates on the radial modes. \par

We have omitted the contributions from micromotion that typically complicate the use of the radial mode. However, recent work by Ratcliffe \textit{et al.}~\cite{Ratcliffe2019} has demonstrated that micromotion can be included in optimisations of the FRAG scheme to enhance fast gate operations. We expect to see a similar enhancement when micromotion is included in optimisations of the GPG and APG schemes. \par

For all global optimisations presented here, the APG and GPG schemes outperform the FRAG and GZC schemes, either in terms of required repetition rate, achievable fidelities, or both. For longer gate times, all four gate schemes are able to find high-fidelity solutions that have only modest repetition rate requirements. In general, it is preferable to use the APG and GPG schemes over FRAG or GZC, as they are compatible with a second phase of optimisation over pulse group timings, which we describe in Section~\ref{sec:local_opt}. \par
\begin{figure}[t!]
    \centering
    \includegraphics[width=0.45\textwidth]{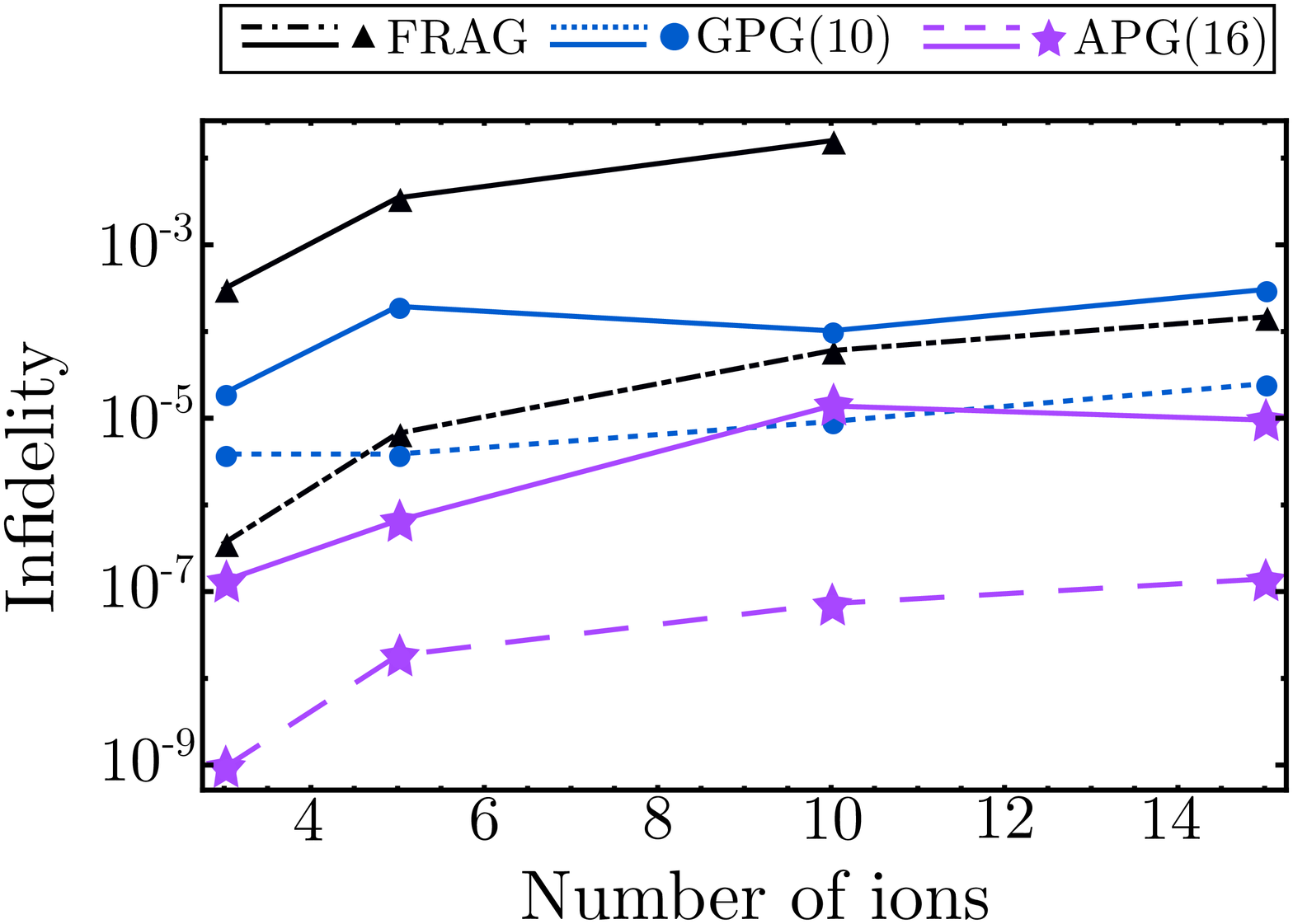}
    \caption{(Colour online). Infidelities of FRAG, GPG(10) and APG(16) gates for increasing numbers of ions in a single Paul trap is increased, for fixed axial frequency of $\omega_t = 2\pi \times \SI{1.2}{\mega\hertz}$. Gates were optimised for each multi-ion system with gate times capped at $T_G = 0.6$ trap periods. The infidelities reported here are for minimum resolving repetition rates of \SI{1}{\giga\hertz} (solid) and \SI{10}{\giga\hertz} (dashed), respectively. The FRAG optimisation for $15$ ions was not able to find gates that are resolvable by \SI{1}{\giga\hertz} repetition rate.}
    \label{fig:LP_scaling}
\end{figure}
\subsection{\label{sec:LP_scaling}Ion scaling in a linear Paul trap }
While gates optimised for a simple two-ion system are robust to the presence of surrounding ions in a microtrap array \cite{Ratcliffe2018}, the same is not true of ion chains in a linear Paul trap except in the limit of infinitely fast gate speeds~\cite{Bentley2015b}. Therefore, it is necessary to include the motional modes of all of the ions in the chain when optimising for a linear Paul trap architecture. Bentley \textit{et al.}~\cite{Bentley2015b} performed this analysis, comparing the FRAG, GZC and Duan schemes, and determined that the FRAG scheme was the best performing. We repeat these calculations here, and compare the FRAG scheme to the GPG$(10)$ and APG$(16)$ schemes. \par

In Fig.~\ref{fig:LP_scaling}, we report infidelities of gates optimised for different numbers of ions in a common Paul trap with fixed axial frequency $\omega_t = 2\pi \times \SI{1.2}{\mega\hertz}$. For a given number of ions, optimisations under the different schemes were each allowed comparable computational time. For all numbers of ions, the GPG and APG optimisations clearly outperform the FRAG scheme, achieving orders of magnitude lower infidelities for given repetition rates. We attribute this to the fact that the FRAG scheme has only three free parameters to optimise over, which is well suited to two-ion systems, where only three conditions need to be satisfied (one phase condition and restoration of two motional modes). As more ions are added, the increased motional modes result in more restoration conditions that need to be satisfied, and thus the FRAG scheme no longer exactly solves Eqs.~\eqref{eq:phase_condition} and \eqref{eq:motional_restoration}. \par

In contrast, the APG and GPG schemes have an increased number of free variables (in this case, eight and ten, respectively) and thus have more freedom to find solutions that satisfy all motional conditions. In particular, the APG scheme appears best suited for multi-ion systems, which is likely due to the anti-symmetric constraints guaranteeing momentum restoration of each motional mode~\eqref{eq:motres_antisym}. The trends in Fig.~\ref{fig:LP_scaling} are not monotonic, which suggests the optimisation is not fully convergent due to an under-sampled search space as the cost function becomes increasingly complex for large numbers of ions. \par
\begin{figure*}[t!]
    \centering
    \includegraphics[width=\linewidth]{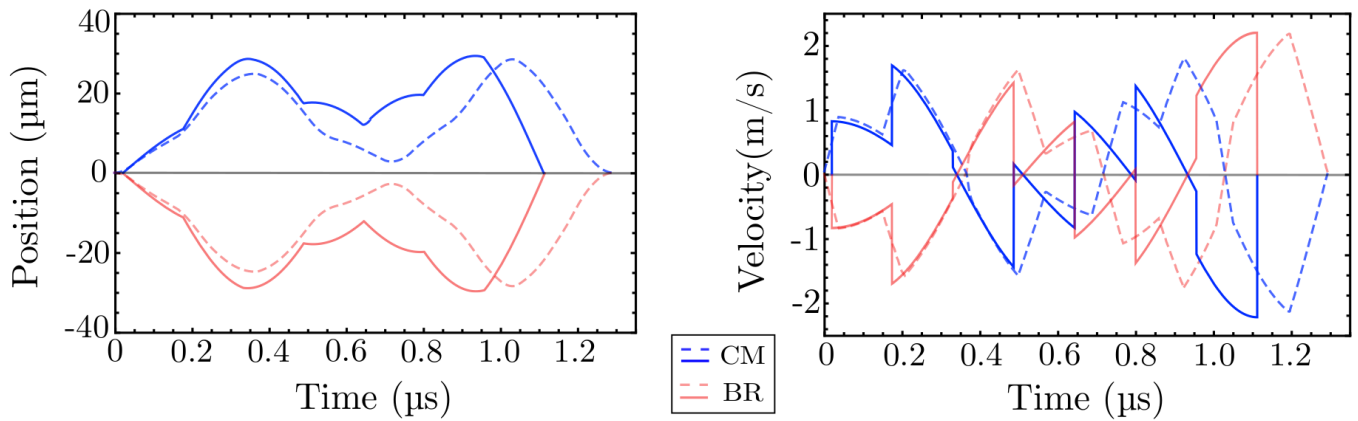} 
    \caption{(Colour online). Motional restoration of a GPG$(8)$ gate for a two-ion microtrap array. The motion of the common-motional (blue) and breathing (red) modes has been plotted by numerically solving the ODEs~\eqref{eq:system_ode}. For visualisation here we assume symmetric trajectories for the qubit states $|\uparrow \uparrow\rangle$ and $|\downarrow\downarrow\rangle$, and the pair $|\uparrow \downarrow\rangle$ and $|\downarrow\uparrow\rangle$. Fidelity calculations take into account asymmetry between these trajectories, as outlined in Appendix \ref{append:ODE}. The solid line represents the motion of a gate solution found from global optimisation, initially with $1 - F = 1.9 \times 10^{-5}$. The dashed lines corresponds to a local optimisation of that gate solution including Coulomb non-linearities and a finite repetition rate of \SI{1}{\giga\hertz}, giving $1 - F = 2.4 \times 10^{-6}$. Motional restoration is clearly demonstrated as both the position and velocity trajectories return to the origin by the end of the gate operation. For effective local re-optimisation, the gate time is allowed to be slightly longer than fixed in global optimisation.}
    \label{fig:nonlinear_effects}
\end{figure*}
\section{\label{sec:local_opt}Local optimisations with ODE description of gate dynamics}
The gate solutions found by the global optimisation phase described in previous sections can be further improved by locally optimising the timings of individual pulse groups. While this can be done inexpensively with the truncated cost function~\eqref{eq:truncatedinf}, it is preferable to utilise an ODE description of the classical motion of the ions for each state-dependent trajectory. This allows for the inclusion of non-linear contributions of the Coulomb interaction, and can be extended to include trap anharmonicities. The ODEs themselves can be derived from the potential energy of the trapped ion system either using Euler-Lagrange equations, or simply from the force equation $-\frac{\partial V}{\partial x_i}=m\ddot{x}_i$. \par

For a two-ion system in a microtrap array with inter-trap distance $d$, this is given by the following pair of coupled differential equations:
\begin{align} \label{eq:system_ode}
\begin{aligned}
    -\frac{e^2}{4 \pi \epsilon_0} \frac{1}{[d + x_1(t) - x_2(t)]^2} + M \omega_t^2 x_1(t) = -M \ddot{x}_1(t) \, , \\
    \frac{e^2}{4 \pi \epsilon_0} \frac{1}{[d + x_1(t) - x_2(t)]^2} + M \omega_t^2 x_2(t) = -M \ddot{x}_2(t) \, ,
\end{aligned}
\end{align}
which describes free evolution of the motional state between state-dependent kicks. Here, $x_i(t)$ is the deviation of the $i$-th ion's position from its equilibrium. In this description, momentum kicks can be modelled as instantaneously transforming the velocities $\dot{x}_i\rightarrow\dot{x}_i\pm z_k(\frac{\hbar k}{M})$, which amount to new initial conditions for integration in Eq.~\eqref{eq:system_ode}. The sign of this transformation depends on the initial two-qubit state. To capture the full dynamics of the gate operation, the above equations must be numerically integrated separately for each initial state. The acquired phase difference can be calculated from the areas enclosed by the state-dependent trajectories through phase space \cite{Garcia-Ripoll2005, Lee2005}, or equivalently from the action of the classical motion of the ions. Calculations of the phase difference and motional restoration are described in Appendix~\ref{append:ODE}. \par

Importantly, the ODE expression can be used to explicitly include the effect of finite repetition rate, by expanding pulse groups into their constituent pairs (as in the Duan scheme \cite{Duan2004a}), and fixing their timings to integer multiples of the laser repetition period, $1/f_{\text{rep}}$. Each pulse pair gives a state-dependent kick of magnitude $2\hbar k$. A lower bound on the infidelity can then be determined by evaluating the ODE for every permutation of the grid-snapped pulse timings. This approach allows for the local optimisation state to be done explicitly for a given laser repetition rate. \par

We exemplify this process using a gate solution found from global optimisation under the GPG$(8)$ scheme, which has a fidelity of $1-2.2\times10^{-6}$ (without considering Coulomb non-linearities). By numerically solving the ODEs~\eqref{eq:system_ode}, we visualise the motional dynamics of the operation in Fig.~\ref{fig:nonlinear_effects}. When the full Coulomb potential is included in the calculation, the fidelity falls to approximately $1-1.9\times10^{-5}$. This drop in fidelity is due to the linearisation around the ions equilibrium position failing when the large state-dependent momentum kicks are applied during the gate operation. When the gate is locally optimised for a finite repetition rate of \SI{1}{\giga\hertz}, the gate fidelity increases to $1-2.4\times10^{-6}$, demonstrating the damage to the fidelity caused by the Coulomb non-linearity can be corrected. In the local re-optimisation we allow the total operation time to be slightly longer than the gate time specified in the global optimisation, which provides more freedom in the timings. We find that allowing the gate time to be a factor of ${\sim}25\%$ longer is an effective constraint on the parameter space, after we do not notice significant improvements in fidelity. For different trapped ion architectures or different gate schemes, this boundary will need to be tuned. \par
\setlength{\tabcolsep}{6pt}
\begin{table*}[t!]
\begin{tabular}{@{}llllllllll@{}}
                     &                    &   &      & $\epsilon=10^{-2}$    & $\epsilon=10^{-3}$ & $\epsilon=10^{-4}$ & $\epsilon=10^{-5}$ & $\epsilon=10^{-6}$ & $\epsilon=10^{-7}$ \\ \hline
                    & $T_G$ & $1-F_0$            & $N$    & \multicolumn{6}{c}{$1-F$}                                                                                                      \\ \hline
                     & $0.45$ & $1.0\times 10^{-4}$          & $1552$ & \multicolumn{1}{c}{-} & $7.0\times 10^{-1}$          & $2.9\times 10^{-1}$          & $3.1\times 10^{-2}$          & $3.2\times 10^{-3}$          & $4.1\times 10^{-4}$          \\
(a)                & $1.0$  & $6.3\times 10^{-9}$          & $640$  & \multicolumn{1}{c}{-} & $8.7\times 10^{-1}$          & $1.2\times 10^{-1}$          & $1.3\times 10^{-2}$          & $1.3\times 10^{-3}$          & $1.3\times 10^{-4}$          \\
                     & $1.75$ & $2.4\times 10^{-7}$          & $191$  & $1.7\times 10^{-1}$             & $3.5\times 10^{-1}$          & $3.8\times 10^{-2}$          & $3.8\times 10^{-3}$          & $3.8\times 10^{-4}$          & $3.8\times 10^{-5}$          \\
                     &                    &        &                       &                    &                    &                    &                    &                    \\
\multicolumn{1}{c}{} & $0.25$ & $1.8\times10^{-4}$ & $1088$ & \multicolumn{1}{c}{-} & $9.9\times 10^{-1}$          & $2.1\times 10^{-1}$          & $2.2\times 10^{-2}$          & $2.4\times 10^{-3}$          & $4.0\times 10^{-4}$          \\
(b)                  & $0.65$ & $3.2\times 10^{-5}$          & $64$   & $8.7\times 10^{-1}$             & $1.2\times 10^{-1}$          & $1.3\times 10^{-2}$          & $1.3\times 10^{-3}$          & $1.6\times 10^{-4}$          & $4.5\times 10^{-5}$          \\
                     & $1.25$ & $2.2\times 10^{-6}$          & $46$   & $7.1\times 10^{-1}$             & $9.0\times 10^{-2}$          & $9.2\times 10^{-3}$          & $9.2\times 10^{-4}$          & $9.4\times 10^{-5}$          & $1.1\times 10^{-5}$          \\ \hline
\end{tabular}
\caption{Realistic infidelities ($1-F$) including worst-case pulse error effects for entangling gates on the longitudinal modes of (a) a two-ion linear microtrap chain ($\chi=1.8\times10^{-4}$), and (b) two-ions in a single linear Paul trap ($\chi=\sqrt{3}-1$). This is reported for a range of gate times $T_G$ (in trap periods), and different values of the transition error $\epsilon$. Here, $F_0$ is the infidelity of the optimised gate not including pulse errors, and $N$ is the number of pulse pairs in the gate. The empty entries correspond to gate errors too large to be described by Eq.~\eqref{eq:fidelity_error}.} 
\label{tab:epsiloninfidelity}
\end{table*}
For a fast gate between two ions in a common Paul trap, the Coulomb non-linearity does not strongly impact gate fidelity, as fewer state-dependent kicks are required to perform a gate between ions that are closer together \cite{Bentley2015b}, and thus the linearisation around the ions' equilibrium positions is typically robust. Consider one of the gate solutions found during global optimisation of the GPG$(10)$ scheme for a gate time of $0.25$ trap periods, chosen from Fig.~\ref{fig:LP2ion}(a), initially with a calculated fidelity of $1-5.6\times10^{-6}$. When the gate scheme is simulated using the ODEs~\eqref{eq:system_ode} (setting $d = 0$), the resulting fidelity calculated with the full Coulomb potential is $1-7.7\times10^{-6}$, which is a much smaller difference than the microtrap example above. \par

\section{\label{sec:pulse_imperfections} Effect of pulse imperfections}
In several previous analyses, it has been identified that errors in pulse area is a limiting factor to experimental implementations of fast gates \cite{Garcia-Ripoll2003, Bentley2016, Taylor2017, Ratcliffe2018}. These errors are typically due to intensity fluctuations of the pulsed laser and lead to individual pulses performing imperfect $\Theta \neq \pi$ rotations on the Bloch sphere. The errors can also arise due to the broadband laser pulses not being resonant with the target atomic transition, which can result from level splitting due to stray magnetic fields or an AC Stark effect. The imperfect pulse areas lead to incorrect rotations of the qubits on the Bloch sphere, which in turn lead to unwanted internal state populations, as well as incorrect motional states that do not disentangle at the end of the gate operation. \par

We present a worst-case analysis of this error, where we suppose each pulse performs an imperfect rotation on the Bloch sphere and investigate systematic effects on gate fidelity. We will assume that each counter-propagating pulse pair is split from a single pulse using a simple pulse-splitting technique \cite{Bentley2013}, and thus the laser phase has no contribution to the gate dynamics. \par

The unitary for a rotation of an imperfect pulse can be written as,
\begin{align}
    \label{forwardpulse_error}
    \hat{U}_k = \left( \hat{\sigma}_{+} e^{-ik\hat{x}} + \hat{\sigma}_{-} e^{ik\hat{x}} \right) \cos{\theta} + i \sin{\theta} \, \hat{\mathfrak{1}} \, ,
\end{align}
where $k$ is the wavenumber of the laser and $\theta$ is a parameterisation of a typical rotation error, with this form chosen to enforce unitarity (ideal case has $\theta = 0$). From this expression, $\sin\theta$ can be understood as a probability amplitude for the part of the unitary that generates an errant orthogonal state to the desired target state. To second order in $\theta$, the unitary for a counter-propagating pulse pair can then be written as
\begin{align}
    \hat{U}_{\text{pair}} &= \hat{U}_{k} \hat{U}_{-k} \\
    &= \left( 1 - \theta^2 \right) \hat{U}_{0} + 2\theta \; \hat{U}_{\text{err}} - \theta^2 \: \hat{\mathfrak{1}} + \mathcal{O}\left(\theta^3\right) \, .
\end{align}
Here, $\hat{U}_{0}$ is the ideal unitary for a state-dependent kick
\begin{align}
    \hat{U}_{0} =
    \begin{pmatrix}
        e^{2ik\hat{x}} & 0 \\ 
        0 & e^{-2ik\hat{x}}
    \end{pmatrix} \, , 
\end{align}
and $\hat{U}_\text{err}$ is an errant unitary, which describes an unrestored internal state and an incorrect motional state,
\begin{align}
    \hat{U}_\text{err} =
    \begin{pmatrix}
        0 & i \cos(k\hat{x}) \\ 
        i \cos(k\hat{x}) & 0
    \end{pmatrix} \, .
\end{align}
While the identity operation $\hat{\mathfrak{1}}$ correctly restores the internal state, the motional state produced is incorrect. The total unitary for a fast gate composed of $N_p$ pulse pairs can be approximated simply by a product of $\hat{U}_\text{pair}$,
\begin{align}
    \label{eq:worstcase_gateunitary}
    \hat{U}_\text{gate} \approx \left( 1 - N_p\theta^2 \right) \; \hat{U}_{\text{N}} + 2N_p\theta \; \hat{U}_{\perp} \, ,
\end{align}
where $\hat{U}_\text{N} \equiv \hat{U}_\text{0}^{N_p}$ and all non-ideal elements have been grouped into $\hat{U}_\perp$. We assume $\hat{U}_\perp$ always produces states orthogonal to those produced by $\hat{U}_\text{N}$. This is a conservative approximation, as it neglects terms that result in an incorrect motional state, but includes those that correctly restore the internal state. In general, these terms will have a small overlap with the ideal final state because the Lamb-Dicke parameter $\eta$ has a finite value. \par

Eq.~\eqref{eq:worstcase_gateunitary} can be used to calculate the effect of $\theta$ on the gate fidelity. For some initial state $|\psi_0\rangle$, the representative-state fidelity becomes
\begin{align}
    F &= \left| \left\langle \psi_0 \left| \hat{U}_{\text{ideal}}^\dag \hat{U}_{\text{gate}} \right| \psi_0 \right\rangle \right|^2 \\
    &= \left| \left( 1-N_p \theta^2 \right) \left\langle \psi_0 \left| \hat{U}_{\text{ideal}}^\dag \hat{U}_\text{N} \right| \psi_0 \right\rangle \right|^2 \, .
\end{align}
Defining $\epsilon \equiv |\theta|^2$ to be the transition error, and $F_0 \equiv |\langle\psi_0|\hat{U}_\text{ideal}^\dag \hat{U}_\text{N} |\psi_0\rangle|^2$ to be the theoretical gate fidelity assuming perfect $\pi$ rotations ($\theta = 0$) the gate fidelity is
\begin{align}
    \label{eq:fidelity_error}
    F \approx \left| 1 - 2 N_p \epsilon + N_p^2 \epsilon^2 \right| F_0 \, .
\end{align}
This expression forms a lower bound on the realistic fidelity for a gate with $N_p$ pulse pairs, and characteristic transition error $\epsilon$. For square pulses, this transition error is determined by the magnitude of relative intensity fluctuations $\Delta I/I$,
\begin{align}
    \label{eq:epsilon_square}
    \epsilon_\text{square} = \frac{\pi^2}{8}\frac{\Delta I}{I} \, .
\end{align}
The fidelity for a range of gate speeds and magnitudes of $\epsilon$ is tabulated in Table \ref{tab:epsiloninfidelity}, for both microtrap and linear Paul trap architectures. Clearly, in order to achieve sub-microsecond gates with fidelities greater than $99.9\%$, transition errors on the order of $10^{-6}$ are required. While single-qubit rotations have been demonstrated with this level of precision \cite{Harty2014}, the current state-of-the-art with ultrafast pulses (tens of picosecond pulses) have only been able to achieve errors of around $10^{-2}$ \cite{Campbell2010, Heinrich2019}. Table \ref{tab:epsiloninfidelity} suggests that if error rates of $10^{-3}$ can be reached, sub-trap period gates between two ions in a Paul trap with fidelities of around $99\%$ can be achieved.  \par

This error can be significantly reduced in future experiments by replacing each individual pulse with a composite pulse sequence that is robust to first or higher-order intensity fluctuations, such as a BB1 pulse sequence \cite{MerrillBrown2014}. Further improvement may be achieved through use of rapid adiabatic passage with chirped laser pulses or simple pulse shaping schemes \cite{Vitanov2001, Goswami2002}. It has previously been stated that the transition error can be reduced by engineering a $\pi$-phase shift between the pulses in each counter-propagating pair to reduce unwanted residual population transfer \cite{Taylor2017, Hussain2016}. However, this generally will not result in any significant improvement, as it does not correct for errant motional states. \par

\section{\label{sec:conclusion}Conclusion}
We have presented a general two-phase approach for optimisation of pulse schemes for implementing fast entangling gate operations on trapped ion platforms. These schemes significantly outperform previous schemes in terms of achievable gate time and/or fidelity. These benefits are particularly important for systems with large numbers of qubits. The scheme design begins with an initial global optimisation phase over the number of pulse pairs in each pulse group and is followed by a second phase of local optimisations over the pulse timings. These local optimisations include the non-linearity of the Coulomb interaction, and explicitly accounts for the finite repetition rate of the laser. Since the local optimisations utilise the full ODE description of the ions' motion, they can be used to include all known measureable quantities of the trapped ion system, such as trap non-linearities. Aspects of the experiment that are not known \textit{a priori} can be accounted for by an online optimisation on the trapped ion machine itself. \par

\begin{acknowledgments}
This research was undertaken with the assistance of resources and services from the National Computational Infrastructure (NCI), which is supported by the Australian Government.
\end{acknowledgments}

\section{Appendices}
\appendix
\section{\label{app:param_choices}Parameter choices}
We consider $^{40}\text{Ca}^{+}$ as the candidate ion species for all calculations presented in this manuscript, with the \SI{393}{\nano\meter} $S_{1/2}\rightarrow P_{3/2}$ transition used for the state-dependent kicks. We use a Lamb-Dicke parameter of $\eta = 0.16$, which corresponds to a trap frequency of $\omega_t \simeq 2\pi \times \SI{1.2}{\mega\hertz}$. Furthermore, we assume an average mode occupation of $\Bar{n} = 0.1$ for each motional mode, which is well within experimental capability \cite{Lechner2016}. By inspection of Eq.~\eqref{eq:truncatedinf}, we can see that as the infidelity is approximately linear with $\Bar{n}$, and because of the very low infidelities we report, significantly higher mode occupations are compatible with high-fidelity gate operations. These parameters are chosen for fair comparison to previous work, namely Ratcliffe \textit{et al.}~\cite{Ratcliffe2018} and Bentley \textit{et al.}~\cite{Bentley2015b}. \par

\section{\label{app:global_opt}Global optimisation methods}
In our approach to global optimisation, we seek to minimise gate infidelity over a parameter space of pulse group timings (FRAG and GZC schemes) or the numbers of pulse pairs in each group (GPG and APG schemes). For the calculations presented in this manuscript, we use the Limited-Memory Broyden-Fletcher-Goldfarb-Shanno (L-BFGS) algorithm \cite{lbfgs_original1989} to minimise the cost-function given in Eq.~\eqref{eq:truncatedinf}, under the assumption of infinite laser repetition rate and a linearised Coulomb potential. Specifically, we apply the L-BFGS-B algorithm \cite{Byrd1995}, where sets of local optimisations are conducted within strict boundaries of the parameter space. In each of these local optimisations the solution with the lowest infidelity is identified as the optimal gate within the restricted region of parameter space. This optimal solution is then chosen as the initial condition for the next local optimisation where the boundaries are slightly widened. This process is repeated until an optimal gate solution is found for a desired range of allowed parameters. For the GPG and APG optimisation procedures, each local optimisation continuously searches over the $z_k$ parameters, which are then rounded to integer values when identifying the optimal solution for a given local optimisation. \par

\section{\label{append:ODE}Infidelity calculation with ODE description}
As described in Section \ref{sec:local_opt}, we perform local optimisations of the pulse timings using an ODE description of the classical state-dependent trajectories of the ions. The ODEs given in Eq.~\eqref{eq:system_ode} are numerically integrated using a fourth-order Runge-Kutta algorithm on XMDS2 \cite{XMDS}. The classical trajectories are simulated for all basis states $\{ |00\rangle, \, |01\rangle, \, |10\rangle, \, |11\rangle \}$ and for an unperturbed trajectory with no state-dependent kicks. It is necessary to integrate trajectories for the $|01\rangle$ and $|10\rangle$ states separately as the non-linearity of the Coulomb potential breaks the symmetry in the ions' motions. \par

To calculate the gate infidelity, we use the following truncated expression
\begin{align}
    \label{eq:truncatedinf_positionbasis}
    1 - F \simeq \frac{2}{3} \Delta \phi^2 + \frac{4}{3} \left( \frac{1}{2} + \Bar{n}_p \right) \left( \Delta P_1^2 + \Delta P_2^2 \right) \, ,
\end{align}
where $\Delta P_i$ is the unrestored motion of the $i$-th ion and $\Delta \phi$ is the phase mismatch. This expression resembles Eq.~\eqref{eq:truncatedinf}, except that we have converted from the mode basis to the position of the individual ions. Eq.~\eqref{eq:truncatedinf_positionbasis} gives an upper-bound on the infidelity (i.e. under-reports achievable fidelity) as the next order terms in the expansion are negative. \par

The motional restoration terms ($\Delta P_i)$ are calculated by subtracting the unperturbed trajectory from each state-dependent trajectory in phase space, and then calculating the non-dimensional distance from the origin:
\begin{align}
\label{eq:ODE_motionalrest}
    \Delta P_i = \sqrt{\frac{M\omega_t}{2\hbar} \Delta x_i(T_G)^2 + \frac{M}{2\hbar\omega_t} \Delta \dot{x}_i(T_G)^2} \, ,
\end{align}
where $\Delta x_i = x_i - x_i^{0}$ and $x_i^{0}$ is the unperturbed trajectory (with no state-dependent kicks) of the $i$-th ion with $T_G$ as the gate time. The position and velocity have been non-dimensionalised by corresponding factors. Each state-dependent trajectory gives a different value of $\Delta P_i$, which we choose for the worst-case when we calculate the infidelity using Eq.~\eqref{eq:truncatedinf_positionbasis}. \par

To calculate the phase mismatch ($\Delta \phi$), we use the classical action as a measure of the phase accumulated along each state-dependent trajectory. This is given by the integral over the classical Lagrangian
\begin{align}
\label{eq:actionphase}
    \Phi_{jk} = \frac{1}{\hbar}\int_{|jk\rangle} \left( \frac{1}{2}M(\dot{x}_{1}^2 + \dot{x}_2^2) - V(x_1, x_2) \right) dt \, ,
\end{align}
where $j, \, k = \{ 0, \, 1 \}$ and $\int_{|jk\rangle}$ denotes an integral over the trajectory of the $|jk\rangle$ state. Ideal phase acquisition, up to a global phase (i.e. one that is common to all $\Phi_{jk}$), can be expressed as:
\begin{align}
\begin{aligned}
    \Phi_{00}^\text{ideal} &= \frac{\pi}{4} \, , & \Phi_{10}^\text{ideal} &= -\frac{\pi}{4} \, , \\
    \Phi_{11}^\text{ideal} &= \frac{\pi}{4} \, , & \Phi_{01}^\text{ideal} &= -\frac{\pi}{4} \, .
\end{aligned}
\end{align}
The phase mismatch $\Delta \phi$ can then be calculated as the worst-case deviation from the ideal case:
\begin{align}
    \Delta \phi = \underset{j, k}{\text{max}} \, \left| \Phi_{jk} - \Phi_{jk}^\text{ideal} \right| \, .
\end{align}
For efficient calculation of the phase accumulated along each state-dependent trajectory, we write Eq.~\eqref{eq:actionphase} as an ODE given by
\begin{align}
    \hbar \, \frac{d\Phi}{dt} = \frac{1}{2}M(\dot{x}_{1}^2 + \dot{x}_2^2) - V(x_1, x_2) \, ,
\end{align}
which we integrate with Eq.~\eqref{eq:system_ode} along each trajectory.

\bibliographystyle{bibsty}
\bibliography{bibliography}

\begin{thebibliography}{58}%
\makeatletter
\providecommand \@ifxundefined [1]{%
 \@ifx{#1\undefined}
}%
\providecommand \@ifnum [1]{%
 \ifnum #1\expandafter \@firstoftwo
 \else \expandafter \@secondoftwo
 \fi
}%
\providecommand \@ifx [1]{%
 \ifx #1\expandafter \@firstoftwo
 \else \expandafter \@secondoftwo
 \fi
}%
\providecommand \natexlab [1]{#1}%
\providecommand \enquote  [1]{``#1''}%
\providecommand \bibnamefont  [1]{#1}%
\providecommand \bibfnamefont [1]{#1}%
\providecommand \citenamefont [1]{#1}%
\providecommand \href@noop [0]{\@secondoftwo}%
\providecommand \href [0]{\begingroup \@sanitize@url \@href}%
\providecommand \@href[1]{\@@startlink{#1}\@@href}%
\providecommand \@@href[1]{\endgroup#1\@@endlink}%
\providecommand \@sanitize@url [0]{\catcode `\\12\catcode `\$12\catcode
  `\&12\catcode `\#12\catcode `\^12\catcode `\_12\catcode `\%12\relax}%
\providecommand \@@startlink[1]{}%
\providecommand \@@endlink[0]{}%
\providecommand \url  [0]{\begingroup\@sanitize@url \@url }%
\providecommand \@url [1]{\endgroup\@href {#1}{\urlprefix }}%
\providecommand \urlprefix  [0]{URL }%
\providecommand \Eprint [0]{\href }%
\providecommand \doibase [0]{http://dx.doi.org/}%
\providecommand \selectlanguage [0]{\@gobble}%
\providecommand \bibinfo  [0]{\@secondoftwo}%
\providecommand \bibfield  [0]{\@secondoftwo}%
\providecommand \translation [1]{[#1]}%
\providecommand \BibitemOpen [0]{}%
\providecommand \bibitemStop [0]{}%
\providecommand \bibitemNoStop [0]{.\EOS\space}%
\providecommand \EOS [0]{\spacefactor3000\relax}%
\providecommand \BibitemShut  [1]{\csname bibitem#1\endcsname}%
\let\auto@bib@innerbib\@empty
\bibitem [{\citenamefont {Feynman}(1982)}]{Feynman1982}%
  \BibitemOpen
  \bibfield  {author} {\bibinfo {author} {\bibfnamefont {R.~P.}\ \bibnamefont
  {Feynman}},\ }\href@noop {} {\bibfield  {journal} {\bibinfo  {journal}
  {International Journal of Theoretical Physics}\ }\textbf {\bibinfo {volume}
  {21}},\ \bibinfo {pages} {467} (\bibinfo {year} {1982})}\BibitemShut
  {NoStop}%
\bibitem [{\citenamefont {Spiller}\ \emph {et~al.}(2005)\citenamefont
  {Spiller}, \citenamefont {Munro}, \citenamefont {Barrett},\ and\
  \citenamefont {Kok}}]{Spiller2005}%
  \BibitemOpen
  \bibfield  {author} {\bibinfo {author} {\bibfnamefont {T.~P.}\ \bibnamefont
  {Spiller}}, \bibinfo {author} {\bibfnamefont {W.~J.}\ \bibnamefont {Munro}},
  \bibinfo {author} {\bibfnamefont {S.~D.}\ \bibnamefont {Barrett}}, \ and\
  \bibinfo {author} {\bibfnamefont {P.}~\bibnamefont {Kok}},\ }\href@noop {}
  {\bibfield  {journal} {\bibinfo  {journal} {Contemporary Physics}\ }\textbf
  {\bibinfo {volume} {46}},\ \bibinfo {pages} {407} (\bibinfo {year}
  {2005})}\BibitemShut {NoStop}%
\bibitem [{\citenamefont {Nielsen}\ and\ \citenamefont
  {Chuang}(2010)}]{Nielsen2010}%
  \BibitemOpen
  \bibfield  {author} {\bibinfo {author} {\bibfnamefont {M.~A.}\ \bibnamefont
  {Nielsen}}\ and\ \bibinfo {author} {\bibfnamefont {I.~L.}\ \bibnamefont
  {Chuang}},\ }\href@noop {} {\emph {\bibinfo {title} {Quantum Computation and
  Quantum Information}}},\ \bibinfo {edition} {2nd}\ ed.\ (\bibinfo
  {publisher} {Cambridge University Press},\ \bibinfo {year}
  {2010})\BibitemShut {NoStop}%
\bibitem [{\citenamefont {Hempel}\ \emph {et~al.}(2018)\citenamefont {Hempel},
  \citenamefont {Maier}, \citenamefont {Romero}, \citenamefont {McClean},
  \citenamefont {Monz}, \citenamefont {Shen}, \citenamefont {Jurcevic},
  \citenamefont {Lanyon}, \citenamefont {Love}, \citenamefont {Babbush},
  \citenamefont {Aspuru-Guzik}, \citenamefont {Blatt},\ and\ \citenamefont
  {Roos}}]{Hempel2018}%
  \BibitemOpen
  \bibfield  {author} {\bibinfo {author} {\bibfnamefont {C.}~\bibnamefont
  {Hempel}}, \bibinfo {author} {\bibfnamefont {C.}~\bibnamefont {Maier}},
  \bibinfo {author} {\bibfnamefont {J.}~\bibnamefont {Romero}}, \bibinfo
  {author} {\bibfnamefont {J.}~\bibnamefont {McClean}}, \bibinfo {author}
  {\bibfnamefont {T.}~\bibnamefont {Monz}}, \bibinfo {author} {\bibfnamefont
  {H.}~\bibnamefont {Shen}}, \bibinfo {author} {\bibfnamefont {P.}~\bibnamefont
  {Jurcevic}}, \bibinfo {author} {\bibfnamefont {B.~P.}\ \bibnamefont
  {Lanyon}}, \bibinfo {author} {\bibfnamefont {P.}~\bibnamefont {Love}},
  \bibinfo {author} {\bibfnamefont {R.}~\bibnamefont {Babbush}}, \bibinfo
  {author} {\bibfnamefont {A.}~\bibnamefont {Aspuru-Guzik}}, \bibinfo {author}
  {\bibfnamefont {R.}~\bibnamefont {Blatt}}, \ and\ \bibinfo {author}
  {\bibfnamefont {C.~F.}\ \bibnamefont {Roos}},\ }\href@noop {} {\bibfield
  {journal} {\bibinfo  {journal} {Phys. Rev. X}\ }\textbf {\bibinfo {volume}
  {8}},\ \bibinfo {pages} {031022} (\bibinfo {year} {2018})}\BibitemShut
  {NoStop}%
\bibitem [{\citenamefont {Lamata}\ \emph {et~al.}(2014)\citenamefont {Lamata},
  \citenamefont {Mezzacapo}, \citenamefont {Casanova},\ and\ \citenamefont
  {Solano}}]{Lamata2014}%
  \BibitemOpen
  \bibfield  {author} {\bibinfo {author} {\bibfnamefont {L.}~\bibnamefont
  {Lamata}}, \bibinfo {author} {\bibfnamefont {A.}~\bibnamefont {Mezzacapo}},
  \bibinfo {author} {\bibfnamefont {J.}~\bibnamefont {Casanova}}, \ and\
  \bibinfo {author} {\bibfnamefont {E.}~\bibnamefont {Solano}},\ }\href@noop {}
  {\bibfield  {journal} {\bibinfo  {journal} {EPJ Quantum Technology}\ }\textbf
  {\bibinfo {volume} {1}},\ \bibinfo {pages} {9} (\bibinfo {year}
  {2014})}\BibitemShut {NoStop}%
\bibitem [{\citenamefont {Ladd}\ \emph {et~al.}(2010)\citenamefont {Ladd},
  \citenamefont {Jelezko}, \citenamefont {Laflamme}, \citenamefont {Nakamura},
  \citenamefont {Monroe},\ and\ \citenamefont {O'Brien}}]{Ladd2010}%
  \BibitemOpen
  \bibfield  {author} {\bibinfo {author} {\bibfnamefont {T.~D.}\ \bibnamefont
  {Ladd}}, \bibinfo {author} {\bibfnamefont {F.}~\bibnamefont {Jelezko}},
  \bibinfo {author} {\bibfnamefont {R.}~\bibnamefont {Laflamme}}, \bibinfo
  {author} {\bibfnamefont {Y.}~\bibnamefont {Nakamura}}, \bibinfo {author}
  {\bibfnamefont {C.}~\bibnamefont {Monroe}}, \ and\ \bibinfo {author}
  {\bibfnamefont {J.~L.}\ \bibnamefont {O'Brien}},\ }\href@noop {} {\bibfield
  {journal} {\bibinfo  {journal} {Nature}\ }\textbf {\bibinfo {volume} {464}},\
  \bibinfo {pages} {45} (\bibinfo {year} {2010})}\BibitemShut {NoStop}%
\bibitem [{\citenamefont {Kok}\ \emph {et~al.}(2007)\citenamefont {Kok},
  \citenamefont {Munro}, \citenamefont {Nemoto}, \citenamefont {Ralph},
  \citenamefont {Dowling},\ and\ \citenamefont {Milburn}}]{Kok2007}%
  \BibitemOpen
  \bibfield  {author} {\bibinfo {author} {\bibfnamefont {P.}~\bibnamefont
  {Kok}}, \bibinfo {author} {\bibfnamefont {W.~J.}\ \bibnamefont {Munro}},
  \bibinfo {author} {\bibfnamefont {K.}~\bibnamefont {Nemoto}}, \bibinfo
  {author} {\bibfnamefont {T.~C.}\ \bibnamefont {Ralph}}, \bibinfo {author}
  {\bibfnamefont {J.~P.}\ \bibnamefont {Dowling}}, \ and\ \bibinfo {author}
  {\bibfnamefont {G.~J.}\ \bibnamefont {Milburn}},\ }\href@noop {} {\bibfield
  {journal} {\bibinfo  {journal} {Rev. Mod. Phys.}\ }\textbf {\bibinfo {volume}
  {79}},\ \bibinfo {pages} {135} (\bibinfo {year} {2007})}\BibitemShut
  {NoStop}%
\bibitem [{\citenamefont {Rong}\ \emph {et~al.}(2017)\citenamefont {Rong},
  \citenamefont {Lu}, \citenamefont {Kong}, \citenamefont {Geng}, \citenamefont
  {Wang}, \citenamefont {Shi}, \citenamefont {Duan},\ and\ \citenamefont
  {Du}}]{Rong2017}%
  \BibitemOpen
  \bibfield  {author} {\bibinfo {author} {\bibfnamefont {X.}~\bibnamefont
  {Rong}}, \bibinfo {author} {\bibfnamefont {D.}~\bibnamefont {Lu}}, \bibinfo
  {author} {\bibfnamefont {X.}~\bibnamefont {Kong}}, \bibinfo {author}
  {\bibfnamefont {J.}~\bibnamefont {Geng}}, \bibinfo {author} {\bibfnamefont
  {Y.}~\bibnamefont {Wang}}, \bibinfo {author} {\bibfnamefont {F.}~\bibnamefont
  {Shi}}, \bibinfo {author} {\bibfnamefont {C.-K.}\ \bibnamefont {Duan}}, \
  and\ \bibinfo {author} {\bibfnamefont {J.}~\bibnamefont {Du}},\ }\href@noop
  {} {\bibfield  {journal} {\bibinfo  {journal} {Advances in Physics: X}\
  }\textbf {\bibinfo {volume} {2}},\ \bibinfo {pages} {125} (\bibinfo {year}
  {2017})}\BibitemShut {NoStop}%
\bibitem [{\citenamefont {Kloeffel}\ and\ \citenamefont
  {Loss}(2013)}]{Kloeffel2013}%
  \BibitemOpen
  \bibfield  {author} {\bibinfo {author} {\bibfnamefont {C.}~\bibnamefont
  {Kloeffel}}\ and\ \bibinfo {author} {\bibfnamefont {D.}~\bibnamefont
  {Loss}},\ }\href@noop {} {\bibfield  {journal} {\bibinfo  {journal} {Annual
  Review of Condensed Matter Physics}\ }\textbf {\bibinfo {volume} {4}},\
  \bibinfo {pages} {51} (\bibinfo {year} {2013})}\BibitemShut {NoStop}%
\bibitem [{\citenamefont {Childress}\ and\ \citenamefont
  {Hanson}(2013)}]{Childress2013}%
  \BibitemOpen
  \bibfield  {author} {\bibinfo {author} {\bibfnamefont {L.}~\bibnamefont
  {Childress}}\ and\ \bibinfo {author} {\bibfnamefont {R.}~\bibnamefont
  {Hanson}},\ }\href@noop {} {\bibfield  {journal} {\bibinfo  {journal} {MRS
  Bulletin}\ }\textbf {\bibinfo {volume} {38}},\ \bibinfo {pages} {134}
  (\bibinfo {year} {2013})}\BibitemShut {NoStop}%
\bibitem [{\citenamefont {Makhlin}\ \emph {et~al.}(2001)\citenamefont
  {Makhlin}, \citenamefont {Sch\"on},\ and\ \citenamefont
  {Shnirman}}]{Makhlin2001}%
  \BibitemOpen
  \bibfield  {author} {\bibinfo {author} {\bibfnamefont {Y.}~\bibnamefont
  {Makhlin}}, \bibinfo {author} {\bibfnamefont {G.}~\bibnamefont {Sch\"on}}, \
  and\ \bibinfo {author} {\bibfnamefont {A.}~\bibnamefont {Shnirman}},\
  }\href@noop {} {\bibfield  {journal} {\bibinfo  {journal} {Rev. Mod. Phys.}\
  }\textbf {\bibinfo {volume} {73}},\ \bibinfo {pages} {357} (\bibinfo {year}
  {2001})}\BibitemShut {NoStop}%
\bibitem [{\citenamefont {H\"{a}ffner}\ \emph {et~al.}(2008)\citenamefont
  {H\"{a}ffner}, \citenamefont {Roos},\ and\ \citenamefont
  {Blatt}}]{Haffner2008}%
  \BibitemOpen
  \bibfield  {author} {\bibinfo {author} {\bibfnamefont {H.}~\bibnamefont
  {H\"{a}ffner}}, \bibinfo {author} {\bibfnamefont {C.~F.}\ \bibnamefont
  {Roos}}, \ and\ \bibinfo {author} {\bibfnamefont {R.}~\bibnamefont {Blatt}},\
  }\href@noop {} {\bibfield  {journal} {\bibinfo  {journal} {Physics Reports}\
  }\textbf {\bibinfo {volume} {469}},\ \bibinfo {pages} {155} (\bibinfo {year}
  {2008})}\BibitemShut {NoStop}%
\bibitem [{\citenamefont {Steane}(1997)}]{Steane1997}%
  \BibitemOpen
  \bibfield  {author} {\bibinfo {author} {\bibfnamefont {A.}~\bibnamefont
  {Steane}},\ }\href@noop {} {\bibfield  {journal} {\bibinfo  {journal}
  {Applied Physics B}\ }\textbf {\bibinfo {volume} {64}},\ \bibinfo {pages}
  {623} (\bibinfo {year} {1997})}\BibitemShut {NoStop}%
\bibitem [{\citenamefont {Wineland}\ \emph {et~al.}(1998)\citenamefont
  {Wineland}, \citenamefont {Monroe}, \citenamefont {Itano}, \citenamefont
  {Leibfried}, \citenamefont {King},\ and\ \citenamefont
  {Meekhof}}]{Wineland1998}%
  \BibitemOpen
  \bibfield  {author} {\bibinfo {author} {\bibfnamefont {D.~J.}\ \bibnamefont
  {Wineland}}, \bibinfo {author} {\bibfnamefont {C.}~\bibnamefont {Monroe}},
  \bibinfo {author} {\bibfnamefont {W.~M.}\ \bibnamefont {Itano}}, \bibinfo
  {author} {\bibfnamefont {D.}~\bibnamefont {Leibfried}}, \bibinfo {author}
  {\bibfnamefont {B.~E.}\ \bibnamefont {King}}, \ and\ \bibinfo {author}
  {\bibfnamefont {D.~M.}\ \bibnamefont {Meekhof}},\ }\href@noop {} {\bibfield
  {journal} {\bibinfo  {journal} {Journal of Research of the National Institute
  of Standards and Technology}\ }\textbf {\bibinfo {volume} {103}},\ \bibinfo
  {pages} {259} (\bibinfo {year} {1998})}\BibitemShut {NoStop}%
\bibitem [{\citenamefont {Nigg}\ \emph {et~al.}(2014)\citenamefont {Nigg},
  \citenamefont {M\"{u}ller}, \citenamefont {Martinez}, \citenamefont
  {Schindler}, \citenamefont {Hennrich}, \citenamefont {Monz}, \citenamefont
  {Martin-Delgado},\ and\ \citenamefont {Blatt}}]{Nigg2014}%
  \BibitemOpen
  \bibfield  {author} {\bibinfo {author} {\bibfnamefont {D.}~\bibnamefont
  {Nigg}}, \bibinfo {author} {\bibfnamefont {M.}~\bibnamefont {M\"{u}ller}},
  \bibinfo {author} {\bibfnamefont {E.~A.}\ \bibnamefont {Martinez}}, \bibinfo
  {author} {\bibfnamefont {P.}~\bibnamefont {Schindler}}, \bibinfo {author}
  {\bibfnamefont {M.}~\bibnamefont {Hennrich}}, \bibinfo {author}
  {\bibfnamefont {T.}~\bibnamefont {Monz}}, \bibinfo {author} {\bibfnamefont
  {M.~A.}\ \bibnamefont {Martin-Delgado}}, \ and\ \bibinfo {author}
  {\bibfnamefont {R.}~\bibnamefont {Blatt}},\ }\href@noop {} {\bibfield
  {journal} {\bibinfo  {journal} {Science}\ }\textbf {\bibinfo {volume}
  {345}},\ \bibinfo {pages} {302} (\bibinfo {year} {2014})}\BibitemShut
  {NoStop}%
\bibitem [{\citenamefont {Friis}\ \emph {et~al.}(2018)\citenamefont {Friis},
  \citenamefont {Marty}, \citenamefont {Maier}, \citenamefont {Hempel},
  \citenamefont {Holz\"apfel}, \citenamefont {Jurcevic}, \citenamefont
  {Plenio}, \citenamefont {Huber}, \citenamefont {Roos}, \citenamefont
  {Blatt},\ and\ \citenamefont {Lanyon}}]{Friis2018}%
  \BibitemOpen
  \bibfield  {author} {\bibinfo {author} {\bibfnamefont {N.}~\bibnamefont
  {Friis}}, \bibinfo {author} {\bibfnamefont {O.}~\bibnamefont {Marty}},
  \bibinfo {author} {\bibfnamefont {C.}~\bibnamefont {Maier}}, \bibinfo
  {author} {\bibfnamefont {C.}~\bibnamefont {Hempel}}, \bibinfo {author}
  {\bibfnamefont {M.}~\bibnamefont {Holz\"apfel}}, \bibinfo {author}
  {\bibfnamefont {P.}~\bibnamefont {Jurcevic}}, \bibinfo {author}
  {\bibfnamefont {M.~B.}\ \bibnamefont {Plenio}}, \bibinfo {author}
  {\bibfnamefont {M.}~\bibnamefont {Huber}}, \bibinfo {author} {\bibfnamefont
  {C.}~\bibnamefont {Roos}}, \bibinfo {author} {\bibfnamefont {R.}~\bibnamefont
  {Blatt}}, \ and\ \bibinfo {author} {\bibfnamefont {B.}~\bibnamefont
  {Lanyon}},\ }\href@noop {} {\bibfield  {journal} {\bibinfo  {journal} {Phys.
  Rev. X}\ }\textbf {\bibinfo {volume} {8}},\ \bibinfo {pages} {021012}
  (\bibinfo {year} {2018})}\BibitemShut {NoStop}%
\bibitem [{\citenamefont {Monroe}\ and\ \citenamefont
  {Kim}(2013)}]{Monroe2013}%
  \BibitemOpen
  \bibfield  {author} {\bibinfo {author} {\bibfnamefont {C.}~\bibnamefont
  {Monroe}}\ and\ \bibinfo {author} {\bibfnamefont {J.}~\bibnamefont {Kim}},\
  }\href@noop {} {\bibfield  {journal} {\bibinfo  {journal} {Science}\ }\textbf
  {\bibinfo {volume} {339}},\ \bibinfo {pages} {1164} (\bibinfo {year}
  {2013})}\BibitemShut {NoStop}%
\bibitem [{\citenamefont {Myerson}\ \emph {et~al.}(2008)\citenamefont
  {Myerson}, \citenamefont {Szwer}, \citenamefont {Webster}, \citenamefont
  {Allcock}, \citenamefont {Curtis}, \citenamefont {Imreh}, \citenamefont
  {Sherman}, \citenamefont {Stacey}, \citenamefont {Steane},\ and\
  \citenamefont {Lucas}}]{Myerson2008}%
  \BibitemOpen
  \bibfield  {author} {\bibinfo {author} {\bibfnamefont {A.~H.}\ \bibnamefont
  {Myerson}}, \bibinfo {author} {\bibfnamefont {D.~J.}\ \bibnamefont {Szwer}},
  \bibinfo {author} {\bibfnamefont {S.~C.}\ \bibnamefont {Webster}}, \bibinfo
  {author} {\bibfnamefont {D.~T.~C.}\ \bibnamefont {Allcock}}, \bibinfo
  {author} {\bibfnamefont {M.~J.}\ \bibnamefont {Curtis}}, \bibinfo {author}
  {\bibfnamefont {G.}~\bibnamefont {Imreh}}, \bibinfo {author} {\bibfnamefont
  {J.~A.}\ \bibnamefont {Sherman}}, \bibinfo {author} {\bibfnamefont {D.~N.}\
  \bibnamefont {Stacey}}, \bibinfo {author} {\bibfnamefont {A.~M.}\
  \bibnamefont {Steane}}, \ and\ \bibinfo {author} {\bibfnamefont {D.~M.}\
  \bibnamefont {Lucas}},\ }\href@noop {} {\bibfield  {journal} {\bibinfo
  {journal} {Phys. Rev. Lett.}\ }\textbf {\bibinfo {volume} {100}},\ \bibinfo
  {pages} {200502} (\bibinfo {year} {2008})}\BibitemShut {NoStop}%
\bibitem [{\citenamefont {Piltz}\ \emph {et~al.}(2014)\citenamefont {Piltz},
  \citenamefont {Sriarunothai}, \citenamefont {Varón},\ and\ \citenamefont
  {Wunderlich}}]{Piltz2014}%
  \BibitemOpen
  \bibfield  {author} {\bibinfo {author} {\bibfnamefont {C.}~\bibnamefont
  {Piltz}}, \bibinfo {author} {\bibfnamefont {T.}~\bibnamefont {Sriarunothai}},
  \bibinfo {author} {\bibfnamefont {A.~F.}\ \bibnamefont {Varón}}, \ and\
  \bibinfo {author} {\bibfnamefont {C.}~\bibnamefont {Wunderlich}},\
  }\href@noop {} {\bibfield  {journal} {\bibinfo  {journal} {Nature
  Communications}\ }\textbf {\bibinfo {volume} {5}},\ \bibinfo {pages} {4679}
  (\bibinfo {year} {2014})}\BibitemShut {NoStop}%
\bibitem [{\citenamefont {Harty}\ \emph {et~al.}(2014)\citenamefont {Harty},
  \citenamefont {Allcock}, \citenamefont {Ballance}, \citenamefont {Guidoni},
  \citenamefont {Janacek}, \citenamefont {Linke}, \citenamefont {Stacey},\ and\
  \citenamefont {Lucas}}]{Harty2014}%
  \BibitemOpen
  \bibfield  {author} {\bibinfo {author} {\bibfnamefont {T.~P.}\ \bibnamefont
  {Harty}}, \bibinfo {author} {\bibfnamefont {D.~T.~C.}\ \bibnamefont
  {Allcock}}, \bibinfo {author} {\bibfnamefont {C.~J.}\ \bibnamefont
  {Ballance}}, \bibinfo {author} {\bibfnamefont {L.}~\bibnamefont {Guidoni}},
  \bibinfo {author} {\bibfnamefont {H.~A.}\ \bibnamefont {Janacek}}, \bibinfo
  {author} {\bibfnamefont {N.~M.}\ \bibnamefont {Linke}}, \bibinfo {author}
  {\bibfnamefont {D.~N.}\ \bibnamefont {Stacey}}, \ and\ \bibinfo {author}
  {\bibfnamefont {D.~M.}\ \bibnamefont {Lucas}},\ }\href@noop {} {\bibfield
  {journal} {\bibinfo  {journal} {Phys. Rev. Lett.}\ }\textbf {\bibinfo
  {volume} {113}},\ \bibinfo {pages} {220501} (\bibinfo {year}
  {2014})}\BibitemShut {NoStop}%
\bibitem [{\citenamefont {Gaebler}\ \emph {et~al.}(2016)\citenamefont
  {Gaebler}, \citenamefont {Tan}, \citenamefont {Lin}, \citenamefont {Wan},
  \citenamefont {Bowler}, \citenamefont {Keith}, \citenamefont {Glancy},
  \citenamefont {Coakley}, \citenamefont {Knill}, \citenamefont {Leibfried},\
  and\ \citenamefont {Wineland}}]{Gaebler2016}%
  \BibitemOpen
  \bibfield  {author} {\bibinfo {author} {\bibfnamefont {J.~P.}\ \bibnamefont
  {Gaebler}}, \bibinfo {author} {\bibfnamefont {T.~R.}\ \bibnamefont {Tan}},
  \bibinfo {author} {\bibfnamefont {Y.}~\bibnamefont {Lin}}, \bibinfo {author}
  {\bibfnamefont {Y.}~\bibnamefont {Wan}}, \bibinfo {author} {\bibfnamefont
  {R.}~\bibnamefont {Bowler}}, \bibinfo {author} {\bibfnamefont {A.~C.}\
  \bibnamefont {Keith}}, \bibinfo {author} {\bibfnamefont {S.}~\bibnamefont
  {Glancy}}, \bibinfo {author} {\bibfnamefont {K.}~\bibnamefont {Coakley}},
  \bibinfo {author} {\bibfnamefont {E.}~\bibnamefont {Knill}}, \bibinfo
  {author} {\bibfnamefont {D.}~\bibnamefont {Leibfried}}, \ and\ \bibinfo
  {author} {\bibfnamefont {D.~J.}\ \bibnamefont {Wineland}},\ }\href@noop {}
  {\bibfield  {journal} {\bibinfo  {journal} {Phys. Rev. Lett.}\ }\textbf
  {\bibinfo {volume} {117}},\ \bibinfo {pages} {060505} (\bibinfo {year}
  {2016})}\BibitemShut {NoStop}%
\bibitem [{\citenamefont {Ballance}\ \emph {et~al.}(2016)\citenamefont
  {Ballance}, \citenamefont {Harty}, \citenamefont {Linke}, \citenamefont
  {Sepiol},\ and\ \citenamefont {Lucas}}]{Ballance2016}%
  \BibitemOpen
  \bibfield  {author} {\bibinfo {author} {\bibfnamefont {C.~J.}\ \bibnamefont
  {Ballance}}, \bibinfo {author} {\bibfnamefont {T.~P.}\ \bibnamefont {Harty}},
  \bibinfo {author} {\bibfnamefont {N.~M.}\ \bibnamefont {Linke}}, \bibinfo
  {author} {\bibfnamefont {M.~A.}\ \bibnamefont {Sepiol}}, \ and\ \bibinfo
  {author} {\bibfnamefont {D.~M.}\ \bibnamefont {Lucas}},\ }\href@noop {}
  {\bibfield  {journal} {\bibinfo  {journal} {Phys. Rev. Lett.}\ }\textbf
  {\bibinfo {volume} {117}},\ \bibinfo {pages} {060504} (\bibinfo {year}
  {2016})}\BibitemShut {NoStop}%
\bibitem [{\citenamefont {Debnath}\ \emph {et~al.}(2016)\citenamefont
  {Debnath}, \citenamefont {Linke}, \citenamefont {Figgatt}, \citenamefont
  {Landsman}, \citenamefont {Wright},\ and\ \citenamefont
  {Monroe}}]{Debnath2016}%
  \BibitemOpen
  \bibfield  {author} {\bibinfo {author} {\bibfnamefont {S.}~\bibnamefont
  {Debnath}}, \bibinfo {author} {\bibfnamefont {N.~M.}\ \bibnamefont {Linke}},
  \bibinfo {author} {\bibfnamefont {C.}~\bibnamefont {Figgatt}}, \bibinfo
  {author} {\bibfnamefont {K.~A.}\ \bibnamefont {Landsman}}, \bibinfo {author}
  {\bibfnamefont {K.}~\bibnamefont {Wright}}, \ and\ \bibinfo {author}
  {\bibfnamefont {C.}~\bibnamefont {Monroe}},\ }\href@noop {} {\bibfield
  {journal} {\bibinfo  {journal} {Nature}\ }\textbf {\bibinfo {volume} {536}},\
  \bibinfo {pages} {63} (\bibinfo {year} {2016})}\BibitemShut {NoStop}%
\bibitem [{\citenamefont {DiVincenzo}(1995)}]{DiVincenzo1995a}%
  \BibitemOpen
  \bibfield  {author} {\bibinfo {author} {\bibfnamefont {D.~P.}\ \bibnamefont
  {DiVincenzo}},\ }\href@noop {} {\bibfield  {journal} {\bibinfo  {journal}
  {Science}\ }\textbf {\bibinfo {volume} {270}},\ \bibinfo {pages} {255}
  (\bibinfo {year} {1995})}\BibitemShut {NoStop}%
\bibitem [{\citenamefont {Ratcliffe}\ \emph {et~al.}(2018)\citenamefont
  {Ratcliffe}, \citenamefont {Taylor}, \citenamefont {Hope},\ and\
  \citenamefont {Carvalho}}]{Ratcliffe2018}%
  \BibitemOpen
  \bibfield  {author} {\bibinfo {author} {\bibfnamefont {A.~K.}\ \bibnamefont
  {Ratcliffe}}, \bibinfo {author} {\bibfnamefont {R.~L.}\ \bibnamefont
  {Taylor}}, \bibinfo {author} {\bibfnamefont {J.~J.}\ \bibnamefont {Hope}}, \
  and\ \bibinfo {author} {\bibfnamefont {A.~R.~R.}\ \bibnamefont {Carvalho}},\
  }\href@noop {} {\bibfield  {journal} {\bibinfo  {journal} {Phys. Rev. Lett.}\
  }\textbf {\bibinfo {volume} {120}},\ \bibinfo {pages} {220501} (\bibinfo
  {year} {2018})}\BibitemShut {NoStop}%
\bibitem [{\citenamefont {Cirac}\ and\ \citenamefont
  {Zoller}(2000)}]{Cirac2000}%
  \BibitemOpen
  \bibfield  {author} {\bibinfo {author} {\bibfnamefont {J.~I.}\ \bibnamefont
  {Cirac}}\ and\ \bibinfo {author} {\bibfnamefont {P.}~\bibnamefont {Zoller}},\
  }\href@noop {} {\bibfield  {journal} {\bibinfo  {journal} {Nature}\ }\textbf
  {\bibinfo {volume} {404}},\ \bibinfo {pages} {579} (\bibinfo {year}
  {2000})}\BibitemShut {NoStop}%
\bibitem [{\citenamefont {Kumph}\ \emph {et~al.}(2016)\citenamefont {Kumph},
  \citenamefont {Holz}, \citenamefont {Langer}, \citenamefont {Meraner},
  \citenamefont {Niedermayr}, \citenamefont {Brownnutt},\ and\ \citenamefont
  {Blatt}}]{Kumph2016}%
  \BibitemOpen
  \bibfield  {author} {\bibinfo {author} {\bibfnamefont {M.}~\bibnamefont
  {Kumph}}, \bibinfo {author} {\bibfnamefont {P.}~\bibnamefont {Holz}},
  \bibinfo {author} {\bibfnamefont {K.}~\bibnamefont {Langer}}, \bibinfo
  {author} {\bibfnamefont {M.}~\bibnamefont {Meraner}}, \bibinfo {author}
  {\bibfnamefont {M.}~\bibnamefont {Niedermayr}}, \bibinfo {author}
  {\bibfnamefont {M.}~\bibnamefont {Brownnutt}}, \ and\ \bibinfo {author}
  {\bibfnamefont {R.}~\bibnamefont {Blatt}},\ }\href@noop {} {\bibfield
  {journal} {\bibinfo  {journal} {New Journal of Physics}\ }\textbf {\bibinfo
  {volume} {18}},\ \bibinfo {pages} {023047} (\bibinfo {year}
  {2016})}\BibitemShut {NoStop}%
\bibitem [{\citenamefont {Cirac}\ and\ \citenamefont
  {Zoller}(1995)}]{Cirac1995}%
  \BibitemOpen
  \bibfield  {author} {\bibinfo {author} {\bibfnamefont {J.~I.}\ \bibnamefont
  {Cirac}}\ and\ \bibinfo {author} {\bibfnamefont {P.}~\bibnamefont {Zoller}},\
  }\href@noop {} {\bibfield  {journal} {\bibinfo  {journal} {Phys. Rev. Lett.}\
  }\textbf {\bibinfo {volume} {74}},\ \bibinfo {pages} {4091} (\bibinfo {year}
  {1995})}\BibitemShut {NoStop}%
\bibitem [{\citenamefont {Jonathan}\ \emph {et~al.}(2000)\citenamefont
  {Jonathan}, \citenamefont {Plenio},\ and\ \citenamefont
  {Knight}}]{Jonathan2000}%
  \BibitemOpen
  \bibfield  {author} {\bibinfo {author} {\bibfnamefont {D.}~\bibnamefont
  {Jonathan}}, \bibinfo {author} {\bibfnamefont {M.~B.}\ \bibnamefont
  {Plenio}}, \ and\ \bibinfo {author} {\bibfnamefont {P.~L.}\ \bibnamefont
  {Knight}},\ }\href@noop {} {\bibfield  {journal} {\bibinfo  {journal} {Phys.
  Rev. A}\ }\textbf {\bibinfo {volume} {62}},\ \bibinfo {pages} {042307}
  (\bibinfo {year} {2000})}\BibitemShut {NoStop}%
\bibitem [{\citenamefont {Schmidt-Kaler}\ \emph
  {et~al.}(2003{\natexlab{a}})\citenamefont {Schmidt-Kaler}, \citenamefont
  {H{\"a}ffner}, \citenamefont {Gulde}, \citenamefont {Riebe}, \citenamefont
  {Lancaster}, \citenamefont {Deuschle}, \citenamefont {Becher}, \citenamefont
  {H{\"a}nsel}, \citenamefont {Eschner}, \citenamefont {Roos},\ and\
  \citenamefont {Blatt}}]{Schmidt-Kaler2003b}%
  \BibitemOpen
  \bibfield  {author} {\bibinfo {author} {\bibfnamefont {F.}~\bibnamefont
  {Schmidt-Kaler}}, \bibinfo {author} {\bibfnamefont {H.}~\bibnamefont
  {H{\"a}ffner}}, \bibinfo {author} {\bibfnamefont {S.}~\bibnamefont {Gulde}},
  \bibinfo {author} {\bibfnamefont {M.}~\bibnamefont {Riebe}}, \bibinfo
  {author} {\bibfnamefont {G.~P.~T.}\ \bibnamefont {Lancaster}}, \bibinfo
  {author} {\bibfnamefont {T.}~\bibnamefont {Deuschle}}, \bibinfo {author}
  {\bibfnamefont {C.}~\bibnamefont {Becher}}, \bibinfo {author} {\bibfnamefont
  {W.}~\bibnamefont {H{\"a}nsel}}, \bibinfo {author} {\bibfnamefont
  {J.}~\bibnamefont {Eschner}}, \bibinfo {author} {\bibfnamefont {C.~F.}\
  \bibnamefont {Roos}}, \ and\ \bibinfo {author} {\bibfnamefont
  {R.}~\bibnamefont {Blatt}},\ }\href@noop {} {\bibfield  {journal} {\bibinfo
  {journal} {Applied Physics B}\ }\textbf {\bibinfo {volume} {77}},\ \bibinfo
  {pages} {789} (\bibinfo {year} {2003}{\natexlab{a}})}\BibitemShut {NoStop}%
\bibitem [{\citenamefont {Schmidt-Kaler}\ \emph
  {et~al.}(2003{\natexlab{b}})\citenamefont {Schmidt-Kaler}, \citenamefont
  {H\"{a}ffner}, \citenamefont {Riebe}, \citenamefont {Gulde}, \citenamefont
  {Lancaster}, \citenamefont {Deuschle}, \citenamefont {Becher}, \citenamefont
  {Roos}, \citenamefont {Eschner},\ and\ \citenamefont
  {Blatt}}]{Schmidt-Kaler2003c}%
  \BibitemOpen
  \bibfield  {author} {\bibinfo {author} {\bibfnamefont {F.}~\bibnamefont
  {Schmidt-Kaler}}, \bibinfo {author} {\bibfnamefont {H.}~\bibnamefont
  {H\"{a}ffner}}, \bibinfo {author} {\bibfnamefont {M.}~\bibnamefont {Riebe}},
  \bibinfo {author} {\bibfnamefont {S.}~\bibnamefont {Gulde}}, \bibinfo
  {author} {\bibfnamefont {G.~P.~T.}\ \bibnamefont {Lancaster}}, \bibinfo
  {author} {\bibfnamefont {T.}~\bibnamefont {Deuschle}}, \bibinfo {author}
  {\bibfnamefont {C.}~\bibnamefont {Becher}}, \bibinfo {author} {\bibfnamefont
  {C.~F.}\ \bibnamefont {Roos}}, \bibinfo {author} {\bibfnamefont
  {J.}~\bibnamefont {Eschner}}, \ and\ \bibinfo {author} {\bibfnamefont
  {R.}~\bibnamefont {Blatt}},\ }\href@noop {} {\bibfield  {journal} {\bibinfo
  {journal} {Nature}\ }\textbf {\bibinfo {volume} {422}},\ \bibinfo {pages}
  {408} (\bibinfo {year} {2003}{\natexlab{b}})}\BibitemShut {NoStop}%
\bibitem [{\citenamefont {M\o{}lmer}\ and\ \citenamefont
  {S\o{}rensen}(1999)}]{Molmer1999}%
  \BibitemOpen
  \bibfield  {author} {\bibinfo {author} {\bibfnamefont {K.}~\bibnamefont
  {M\o{}lmer}}\ and\ \bibinfo {author} {\bibfnamefont {A.}~\bibnamefont
  {S\o{}rensen}},\ }\href@noop {} {\bibfield  {journal} {\bibinfo  {journal}
  {Phys. Rev. Lett.}\ }\textbf {\bibinfo {volume} {82}},\ \bibinfo {pages}
  {1835} (\bibinfo {year} {1999})}\BibitemShut {NoStop}%
\bibitem [{\citenamefont {S\o{}rensen}\ and\ \citenamefont
  {M\o{}lmer}(1999)}]{Sorensen1999}%
  \BibitemOpen
  \bibfield  {author} {\bibinfo {author} {\bibfnamefont {A.}~\bibnamefont
  {S\o{}rensen}}\ and\ \bibinfo {author} {\bibfnamefont {K.}~\bibnamefont
  {M\o{}lmer}},\ }\href@noop {} {\bibfield  {journal} {\bibinfo  {journal}
  {Phys. Rev. Lett.}\ }\textbf {\bibinfo {volume} {82}},\ \bibinfo {pages}
  {1971} (\bibinfo {year} {1999})}\BibitemShut {NoStop}%
\bibitem [{\citenamefont {S\o{}rensen}\ and\ \citenamefont
  {M\o{}lmer}(2000)}]{Sorensen2000}%
  \BibitemOpen
  \bibfield  {author} {\bibinfo {author} {\bibfnamefont {A.}~\bibnamefont
  {S\o{}rensen}}\ and\ \bibinfo {author} {\bibfnamefont {K.}~\bibnamefont
  {M\o{}lmer}},\ }\href@noop {} {\bibfield  {journal} {\bibinfo  {journal}
  {Phys. Rev. A}\ }\textbf {\bibinfo {volume} {62}},\ \bibinfo {pages} {022311}
  (\bibinfo {year} {2000})}\BibitemShut {NoStop}%
\bibitem [{\citenamefont {Zhu}\ \emph {et~al.}(2006)\citenamefont {Zhu},
  \citenamefont {Monroe},\ and\ \citenamefont {Duan}}]{Zhu2006}%
  \BibitemOpen
  \bibfield  {author} {\bibinfo {author} {\bibfnamefont {S.-L.}\ \bibnamefont
  {Zhu}}, \bibinfo {author} {\bibfnamefont {C.}~\bibnamefont {Monroe}}, \ and\
  \bibinfo {author} {\bibfnamefont {L.-M.}\ \bibnamefont {Duan}},\ }\href@noop
  {} {\bibfield  {journal} {\bibinfo  {journal} {EPL (Europhysics Letters)}\
  }\textbf {\bibinfo {volume} {73}},\ \bibinfo {pages} {485} (\bibinfo {year}
  {2006})}\BibitemShut {NoStop}%
\bibitem [{\citenamefont {Garc\'{i}a-Ripoll}\ \emph {et~al.}(2003)\citenamefont
  {Garc\'{i}a-Ripoll}, \citenamefont {Zoller},\ and\ \citenamefont
  {Cirac}}]{Garcia-Ripoll2003}%
  \BibitemOpen
  \bibfield  {author} {\bibinfo {author} {\bibfnamefont {J.~J.}\ \bibnamefont
  {Garc\'{i}a-Ripoll}}, \bibinfo {author} {\bibfnamefont {P.}~\bibnamefont
  {Zoller}}, \ and\ \bibinfo {author} {\bibfnamefont {J.~I.}\ \bibnamefont
  {Cirac}},\ }\href@noop {} {\bibfield  {journal} {\bibinfo  {journal} {Phys.
  Rev. Lett.}\ }\textbf {\bibinfo {volume} {91}},\ \bibinfo {pages} {157901}
  (\bibinfo {year} {2003})}\BibitemShut {NoStop}%
\bibitem [{\citenamefont {Garc\'{i}a-Ripoll}\ \emph {et~al.}(2005)\citenamefont
  {Garc\'{i}a-Ripoll}, \citenamefont {Zoller},\ and\ \citenamefont
  {Cirac}}]{Garcia-Ripoll2005}%
  \BibitemOpen
  \bibfield  {author} {\bibinfo {author} {\bibfnamefont {J.~J.}\ \bibnamefont
  {Garc\'{i}a-Ripoll}}, \bibinfo {author} {\bibfnamefont {P.}~\bibnamefont
  {Zoller}}, \ and\ \bibinfo {author} {\bibfnamefont {J.~I.}\ \bibnamefont
  {Cirac}},\ }\href@noop {} {\bibfield  {journal} {\bibinfo  {journal} {Phys.
  Rev. A}\ }\textbf {\bibinfo {volume} {71}},\ \bibinfo {pages} {062309}
  (\bibinfo {year} {2005})}\BibitemShut {NoStop}%
\bibitem [{\citenamefont {Sch\"{a}fer}\ \emph {et~al.}(2018)\citenamefont
  {Sch\"{a}fer}, \citenamefont {Ballance}, \citenamefont {Thirumalai},
  \citenamefont {Stephenson}, \citenamefont {Ballance}, \citenamefont
  {Steane},\ and\ \citenamefont {Lucas}}]{Schafer2018}%
  \BibitemOpen
  \bibfield  {author} {\bibinfo {author} {\bibfnamefont {V.~M.}\ \bibnamefont
  {Sch\"{a}fer}}, \bibinfo {author} {\bibfnamefont {C.~J.}\ \bibnamefont
  {Ballance}}, \bibinfo {author} {\bibfnamefont {K.}~\bibnamefont
  {Thirumalai}}, \bibinfo {author} {\bibfnamefont {L.~J.}\ \bibnamefont
  {Stephenson}}, \bibinfo {author} {\bibfnamefont {T.~G.}\ \bibnamefont
  {Ballance}}, \bibinfo {author} {\bibfnamefont {A.~M.}\ \bibnamefont
  {Steane}}, \ and\ \bibinfo {author} {\bibfnamefont {D.~M.}\ \bibnamefont
  {Lucas}},\ }\href@noop {} {\bibfield  {journal} {\bibinfo  {journal}
  {Nature}\ }\textbf {\bibinfo {volume} {555}},\ \bibinfo {pages} {75}
  (\bibinfo {year} {2018})}\BibitemShut {NoStop}%
\bibitem [{\citenamefont {Steane}\ \emph {et~al.}(2014)\citenamefont {Steane},
  \citenamefont {Imreh}, \citenamefont {Home},\ and\ \citenamefont
  {Leibfried}}]{Steane2014}%
  \BibitemOpen
  \bibfield  {author} {\bibinfo {author} {\bibfnamefont {A.~M.}\ \bibnamefont
  {Steane}}, \bibinfo {author} {\bibfnamefont {G.}~\bibnamefont {Imreh}},
  \bibinfo {author} {\bibfnamefont {J.~P.}\ \bibnamefont {Home}}, \ and\
  \bibinfo {author} {\bibfnamefont {D.}~\bibnamefont {Leibfried}},\ }\href@noop
  {} {\bibfield  {journal} {\bibinfo  {journal} {New Journal of Physics}\
  }\textbf {\bibinfo {volume} {16}},\ \bibinfo {pages} {053049} (\bibinfo
  {year} {2014})}\BibitemShut {NoStop}%
\bibitem [{\citenamefont {Duan}(2004)}]{Duan2004a}%
  \BibitemOpen
  \bibfield  {author} {\bibinfo {author} {\bibfnamefont {L.-M.}\ \bibnamefont
  {Duan}},\ }\href@noop {} {\bibfield  {journal} {\bibinfo  {journal} {Phys.
  Rev. Lett.}\ }\textbf {\bibinfo {volume} {93}},\ \bibinfo {pages} {100502}
  (\bibinfo {year} {2004})}\BibitemShut {NoStop}%
\bibitem [{\citenamefont {Bentley}\ \emph {et~al.}(2013)\citenamefont
  {Bentley}, \citenamefont {Carvalho}, \citenamefont {Kielpinski},\ and\
  \citenamefont {Hope}}]{Bentley2013}%
  \BibitemOpen
  \bibfield  {author} {\bibinfo {author} {\bibfnamefont {C.~D.~B.}\
  \bibnamefont {Bentley}}, \bibinfo {author} {\bibfnamefont {A.~R.~R.}\
  \bibnamefont {Carvalho}}, \bibinfo {author} {\bibfnamefont {D.}~\bibnamefont
  {Kielpinski}}, \ and\ \bibinfo {author} {\bibfnamefont {J.~J.}\ \bibnamefont
  {Hope}},\ }\href@noop {} {\bibfield  {journal} {\bibinfo  {journal} {New
  Journal of Physics}\ }\textbf {\bibinfo {volume} {15}},\ \bibinfo {pages}
  {043006} (\bibinfo {year} {2013})}\BibitemShut {NoStop}%
\bibitem [{\citenamefont {Bentley}\ \emph {et~al.}(2015)\citenamefont
  {Bentley}, \citenamefont {Carvalho},\ and\ \citenamefont
  {Hope}}]{Bentley2015b}%
  \BibitemOpen
  \bibfield  {author} {\bibinfo {author} {\bibfnamefont {C.~D.~B.}\
  \bibnamefont {Bentley}}, \bibinfo {author} {\bibfnamefont {A.~R.~R.}\
  \bibnamefont {Carvalho}}, \ and\ \bibinfo {author} {\bibfnamefont {J.~J.}\
  \bibnamefont {Hope}},\ }\href@noop {} {\bibfield  {journal} {\bibinfo
  {journal} {New Journal of Physics}\ }\textbf {\bibinfo {volume} {17}},\
  \bibinfo {pages} {103025} (\bibinfo {year} {2015})}\BibitemShut {NoStop}%
\bibitem [{\citenamefont {Bentley}\ \emph {et~al.}(2016)\citenamefont
  {Bentley}, \citenamefont {Taylor}, \citenamefont {Carvalho},\ and\
  \citenamefont {Hope}}]{Bentley2016}%
  \BibitemOpen
  \bibfield  {author} {\bibinfo {author} {\bibfnamefont {C.~D.~B.}\
  \bibnamefont {Bentley}}, \bibinfo {author} {\bibfnamefont {R.~L.}\
  \bibnamefont {Taylor}}, \bibinfo {author} {\bibfnamefont {A.~R.~R.}\
  \bibnamefont {Carvalho}}, \ and\ \bibinfo {author} {\bibfnamefont {J.~J.}\
  \bibnamefont {Hope}},\ }\href@noop {} {\bibfield  {journal} {\bibinfo
  {journal} {Phys. Rev. A}\ }\textbf {\bibinfo {volume} {93}},\ \bibinfo
  {pages} {042342} (\bibinfo {year} {2016})}\BibitemShut {NoStop}%
\bibitem [{\citenamefont {Hussain}\ \emph {et~al.}(2016)\citenamefont
  {Hussain}, \citenamefont {Petrasiunas}, \citenamefont {Bentley},
  \citenamefont {Taylor}, \citenamefont {Carvalho}, \citenamefont {Hope},
  \citenamefont {Streed}, \citenamefont {Lobino},\ and\ \citenamefont
  {Kielpinski}}]{Hussain2016}%
  \BibitemOpen
  \bibfield  {author} {\bibinfo {author} {\bibfnamefont {M.~I.}\ \bibnamefont
  {Hussain}}, \bibinfo {author} {\bibfnamefont {M.~J.}\ \bibnamefont
  {Petrasiunas}}, \bibinfo {author} {\bibfnamefont {C.~D.~B.}\ \bibnamefont
  {Bentley}}, \bibinfo {author} {\bibfnamefont {R.~L.}\ \bibnamefont {Taylor}},
  \bibinfo {author} {\bibfnamefont {A.~R.~R.}\ \bibnamefont {Carvalho}},
  \bibinfo {author} {\bibfnamefont {J.~J.}\ \bibnamefont {Hope}}, \bibinfo
  {author} {\bibfnamefont {E.~W.}\ \bibnamefont {Streed}}, \bibinfo {author}
  {\bibfnamefont {M.}~\bibnamefont {Lobino}}, \ and\ \bibinfo {author}
  {\bibfnamefont {D.}~\bibnamefont {Kielpinski}},\ }\href@noop {} {\bibfield
  {journal} {\bibinfo  {journal} {Opt. Express}\ }\textbf {\bibinfo {volume}
  {24}},\ \bibinfo {pages} {16638} (\bibinfo {year} {2016})}\BibitemShut
  {NoStop}%
\bibitem [{\citenamefont {Heinrich}\ \emph {et~al.}(2019)\citenamefont
  {Heinrich}, \citenamefont {Guggemos}, \citenamefont {Guevara-Bertsch},
  \citenamefont {Hussain}, \citenamefont {Roos},\ and\ \citenamefont
  {Blatt}}]{Heinrich2019}%
  \BibitemOpen
  \bibfield  {author} {\bibinfo {author} {\bibfnamefont {D.}~\bibnamefont
  {Heinrich}}, \bibinfo {author} {\bibfnamefont {M.}~\bibnamefont {Guggemos}},
  \bibinfo {author} {\bibfnamefont {M.}~\bibnamefont {Guevara-Bertsch}},
  \bibinfo {author} {\bibfnamefont {M.~I.}\ \bibnamefont {Hussain}}, \bibinfo
  {author} {\bibfnamefont {C.~F.}\ \bibnamefont {Roos}}, \ and\ \bibinfo
  {author} {\bibfnamefont {R.}~\bibnamefont {Blatt}},\ }\href@noop {}
  {\bibfield  {journal} {\bibinfo  {journal} {New Journal of Physics}\ }\textbf
  {\bibinfo {volume} {21}},\ \bibinfo {pages} {073017} (\bibinfo {year}
  {2019})}\BibitemShut {NoStop}%
\bibitem [{\citenamefont {Glaser}\ \emph {et~al.}(2015)\citenamefont {Glaser},
  \citenamefont {Boscain}, \citenamefont {Calarco}, \citenamefont {Koch},
  \citenamefont {K{\"o}ckenberger}, \citenamefont {Kosloff}, \citenamefont
  {Kuprov}, \citenamefont {Luy}, \citenamefont {Schirmer}, \citenamefont
  {Schulte-Herbr{\"u}ggen}, \citenamefont {Sugny},\ and\ \citenamefont
  {Wilhelm}}]{Glaser2015}%
  \BibitemOpen
  \bibfield  {author} {\bibinfo {author} {\bibfnamefont {S.~J.}\ \bibnamefont
  {Glaser}}, \bibinfo {author} {\bibfnamefont {U.}~\bibnamefont {Boscain}},
  \bibinfo {author} {\bibfnamefont {T.}~\bibnamefont {Calarco}}, \bibinfo
  {author} {\bibfnamefont {C.~P.}\ \bibnamefont {Koch}}, \bibinfo {author}
  {\bibfnamefont {W.}~\bibnamefont {K{\"o}ckenberger}}, \bibinfo {author}
  {\bibfnamefont {R.}~\bibnamefont {Kosloff}}, \bibinfo {author} {\bibfnamefont
  {I.}~\bibnamefont {Kuprov}}, \bibinfo {author} {\bibfnamefont
  {B.}~\bibnamefont {Luy}}, \bibinfo {author} {\bibfnamefont {S.}~\bibnamefont
  {Schirmer}}, \bibinfo {author} {\bibfnamefont {T.}~\bibnamefont
  {Schulte-Herbr{\"u}ggen}}, \bibinfo {author} {\bibfnamefont {D.}~\bibnamefont
  {Sugny}}, \ and\ \bibinfo {author} {\bibfnamefont {F.~K.}\ \bibnamefont
  {Wilhelm}},\ }\href@noop {} {\bibfield  {journal} {\bibinfo  {journal} {The
  European Physical Journal D}\ }\textbf {\bibinfo {volume} {69}},\ \bibinfo
  {pages} {279} (\bibinfo {year} {2015})}\BibitemShut {NoStop}%
\bibitem [{\citenamefont {Taylor}\ \emph {et~al.}(2017)\citenamefont {Taylor},
  \citenamefont {Bentley}, \citenamefont {Pedernales}, \citenamefont {Lamata},
  \citenamefont {Solano}, \citenamefont {Carvalho},\ and\ \citenamefont
  {Hope}}]{Taylor2017}%
  \BibitemOpen
  \bibfield  {author} {\bibinfo {author} {\bibfnamefont {R.~L.}\ \bibnamefont
  {Taylor}}, \bibinfo {author} {\bibfnamefont {C.~D.~B.}\ \bibnamefont
  {Bentley}}, \bibinfo {author} {\bibfnamefont {J.~S.}\ \bibnamefont
  {Pedernales}}, \bibinfo {author} {\bibfnamefont {L.}~\bibnamefont {Lamata}},
  \bibinfo {author} {\bibfnamefont {E.}~\bibnamefont {Solano}}, \bibinfo
  {author} {\bibfnamefont {A.~R.~R.}\ \bibnamefont {Carvalho}}, \ and\ \bibinfo
  {author} {\bibfnamefont {J.~J.}\ \bibnamefont {Hope}},\ }\href@noop {}
  {\bibfield  {journal} {\bibinfo  {journal} {Scientific Reports}\ }\textbf
  {\bibinfo {volume} {7}},\ \bibinfo {pages} {46197} (\bibinfo {year}
  {2017})}\BibitemShut {NoStop}%
\bibitem [{\citenamefont {Pedersen}\ \emph {et~al.}(2008)\citenamefont
  {Pedersen}, \citenamefont {M\o{}ller},\ and\ \citenamefont
  {M\o{}lmer}}]{Pedersen2008}%
  \BibitemOpen
  \bibfield  {author} {\bibinfo {author} {\bibfnamefont {L.~H.}\ \bibnamefont
  {Pedersen}}, \bibinfo {author} {\bibfnamefont {N.~M.}\ \bibnamefont
  {M\o{}ller}}, \ and\ \bibinfo {author} {\bibfnamefont {K.}~\bibnamefont
  {M\o{}lmer}},\ }\href@noop {} {\bibfield  {journal} {\bibinfo  {journal}
  {Phys. Lett. A}\ }\textbf {\bibinfo {volume} {372}},\ \bibinfo {pages} {7028
  } (\bibinfo {year} {2008})}\BibitemShut {NoStop}%
\bibitem [{\citenamefont {Ratcliffe}\ \emph {et~al.}(2019)\citenamefont
  {Ratcliffe}, \citenamefont {Oberg},\ and\ \citenamefont
  {Hope}}]{Ratcliffe2019}%
  \BibitemOpen
  \bibfield  {author} {\bibinfo {author} {\bibfnamefont {A.~K.}\ \bibnamefont
  {Ratcliffe}}, \bibinfo {author} {\bibfnamefont {L.~M.}\ \bibnamefont
  {Oberg}}, \ and\ \bibinfo {author} {\bibfnamefont {J.~J.}\ \bibnamefont
  {Hope}},\ }\href@noop {} {\enquote {\bibinfo {title} {Micromotion-enhanced
  fast entangling gates for trapped ion quantum computing},}\ } (\bibinfo
  {year} {2019}),\ \Eprint {http://arxiv.org/abs/1902.06364} {arXiv:1902.06364
  [quant-ph]} \BibitemShut {NoStop}%
\bibitem [{\citenamefont {Lee}\ \emph {et~al.}(2005)\citenamefont {Lee},
  \citenamefont {Brickman}, \citenamefont {Deslauriers}, \citenamefont
  {Haljan}, \citenamefont {Duan},\ and\ \citenamefont {Monroe}}]{Lee2005}%
  \BibitemOpen
  \bibfield  {author} {\bibinfo {author} {\bibfnamefont {P.~J.}\ \bibnamefont
  {Lee}}, \bibinfo {author} {\bibfnamefont {K.-A.}\ \bibnamefont {Brickman}},
  \bibinfo {author} {\bibfnamefont {L.}~\bibnamefont {Deslauriers}}, \bibinfo
  {author} {\bibfnamefont {P.~C.}\ \bibnamefont {Haljan}}, \bibinfo {author}
  {\bibfnamefont {L.-M.}\ \bibnamefont {Duan}}, \ and\ \bibinfo {author}
  {\bibfnamefont {C.}~\bibnamefont {Monroe}},\ }\href@noop {} {\bibfield
  {journal} {\bibinfo  {journal} {Journal of Optics B: Quantum and
  Semiclassical Optics}\ }\textbf {\bibinfo {volume} {7}},\ \bibinfo {pages}
  {S371} (\bibinfo {year} {2005})}\BibitemShut {NoStop}%
\bibitem [{\citenamefont {Campbell}\ \emph {et~al.}(2010)\citenamefont
  {Campbell}, \citenamefont {Mizrahi}, \citenamefont {Quraishi}, \citenamefont
  {Senko}, \citenamefont {Hayes}, \citenamefont {Hucul}, \citenamefont
  {Matsukevich}, \citenamefont {Maunz},\ and\ \citenamefont
  {Monroe}}]{Campbell2010}%
  \BibitemOpen
  \bibfield  {author} {\bibinfo {author} {\bibfnamefont {W.~C.}\ \bibnamefont
  {Campbell}}, \bibinfo {author} {\bibfnamefont {J.}~\bibnamefont {Mizrahi}},
  \bibinfo {author} {\bibfnamefont {Q.}~\bibnamefont {Quraishi}}, \bibinfo
  {author} {\bibfnamefont {C.}~\bibnamefont {Senko}}, \bibinfo {author}
  {\bibfnamefont {D.}~\bibnamefont {Hayes}}, \bibinfo {author} {\bibfnamefont
  {D.}~\bibnamefont {Hucul}}, \bibinfo {author} {\bibfnamefont {D.~N.}\
  \bibnamefont {Matsukevich}}, \bibinfo {author} {\bibfnamefont
  {P.}~\bibnamefont {Maunz}}, \ and\ \bibinfo {author} {\bibfnamefont
  {C.}~\bibnamefont {Monroe}},\ }\href@noop {} {\bibfield  {journal} {\bibinfo
  {journal} {Phys. Rev. Lett.}\ }\textbf {\bibinfo {volume} {105}},\ \bibinfo
  {pages} {090502} (\bibinfo {year} {2010})}\BibitemShut {NoStop}%
\bibitem [{\citenamefont {Merrill}\ and\ \citenamefont
  {Brown}(2014)}]{MerrillBrown2014}%
  \BibitemOpen
  \bibfield  {author} {\bibinfo {author} {\bibfnamefont {J.~T.}\ \bibnamefont
  {Merrill}}\ and\ \bibinfo {author} {\bibfnamefont {K.~R.}\ \bibnamefont
  {Brown}},\ }\href@noop {} {\emph {\bibinfo {title} {Quantum Information and
  Computation for Chemistry}}}\ (\bibinfo  {publisher} {John Wiley \& Sons,
  Ltd},\ \bibinfo {year} {2014})\ Chap.~\bibinfo {chapter} {10}, pp.\ \bibinfo
  {pages} {241--294}\BibitemShut {NoStop}%
\bibitem [{\citenamefont {Vitanov}\ \emph {et~al.}(2001)\citenamefont
  {Vitanov}, \citenamefont {Halfmann}, \citenamefont {Shore},\ and\
  \citenamefont {Bergmann}}]{Vitanov2001}%
  \BibitemOpen
  \bibfield  {author} {\bibinfo {author} {\bibfnamefont {N.~V.}\ \bibnamefont
  {Vitanov}}, \bibinfo {author} {\bibfnamefont {T.}~\bibnamefont {Halfmann}},
  \bibinfo {author} {\bibfnamefont {B.~W.}\ \bibnamefont {Shore}}, \ and\
  \bibinfo {author} {\bibfnamefont {K.}~\bibnamefont {Bergmann}},\ }\href@noop
  {} {\bibfield  {journal} {\bibinfo  {journal} {Annu. Rev. Phys. Chem.}\
  }\textbf {\bibinfo {volume} {52}},\ \bibinfo {pages} {763} (\bibinfo {year}
  {2001})}\BibitemShut {NoStop}%
\bibitem [{\citenamefont {Sharan}\ and\ \citenamefont
  {Goswami}(2002)}]{Goswami2002}%
  \BibitemOpen
  \bibfield  {author} {\bibinfo {author} {\bibfnamefont {A.}~\bibnamefont
  {Sharan}}\ and\ \bibinfo {author} {\bibfnamefont {D.}~\bibnamefont
  {Goswami}},\ }\href@noop {} {\bibfield  {journal} {\bibinfo  {journal}
  {Current Science}\ }\textbf {\bibinfo {volume} {82}},\ \bibinfo {pages} {30}
  (\bibinfo {year} {2002})}\BibitemShut {NoStop}%
\bibitem [{\citenamefont {Lechner}\ \emph {et~al.}(2016)\citenamefont
  {Lechner}, \citenamefont {Maier}, \citenamefont {Hempel}, \citenamefont
  {Jurcevic}, \citenamefont {Lanyon}, \citenamefont {Monz}, \citenamefont
  {Brownnutt}, \citenamefont {Blatt},\ and\ \citenamefont
  {Roos}}]{Lechner2016}%
  \BibitemOpen
  \bibfield  {author} {\bibinfo {author} {\bibfnamefont {R.}~\bibnamefont
  {Lechner}}, \bibinfo {author} {\bibfnamefont {C.}~\bibnamefont {Maier}},
  \bibinfo {author} {\bibfnamefont {C.}~\bibnamefont {Hempel}}, \bibinfo
  {author} {\bibfnamefont {P.}~\bibnamefont {Jurcevic}}, \bibinfo {author}
  {\bibfnamefont {B.~P.}\ \bibnamefont {Lanyon}}, \bibinfo {author}
  {\bibfnamefont {T.}~\bibnamefont {Monz}}, \bibinfo {author} {\bibfnamefont
  {M.}~\bibnamefont {Brownnutt}}, \bibinfo {author} {\bibfnamefont
  {R.}~\bibnamefont {Blatt}}, \ and\ \bibinfo {author} {\bibfnamefont {C.~F.}\
  \bibnamefont {Roos}},\ }\href@noop {} {\bibfield  {journal} {\bibinfo
  {journal} {Phys. Rev. A}\ }\textbf {\bibinfo {volume} {93}},\ \bibinfo
  {pages} {053401} (\bibinfo {year} {2016})}\BibitemShut {NoStop}%
\bibitem [{\citenamefont {Liu}\ and\ \citenamefont
  {Nocedal}(1989)}]{lbfgs_original1989}%
  \BibitemOpen
  \bibfield  {author} {\bibinfo {author} {\bibfnamefont {D.~C.}\ \bibnamefont
  {Liu}}\ and\ \bibinfo {author} {\bibfnamefont {J.}~\bibnamefont {Nocedal}},\
  }\href@noop {} {\bibfield  {journal} {\bibinfo  {journal} {Mathematical
  Programming}\ }\textbf {\bibinfo {volume} {45}},\ \bibinfo {pages}
  {503–528} (\bibinfo {year} {1989})}\BibitemShut {NoStop}%
\bibitem [{\citenamefont {Byrd}\ \emph {et~al.}(1995)\citenamefont {Byrd},
  \citenamefont {Lu}, \citenamefont {Nocedal},\ and\ \citenamefont
  {Zhu}}]{Byrd1995}%
  \BibitemOpen
  \bibfield  {author} {\bibinfo {author} {\bibfnamefont {R.~H.}\ \bibnamefont
  {Byrd}}, \bibinfo {author} {\bibfnamefont {P.}~\bibnamefont {Lu}}, \bibinfo
  {author} {\bibfnamefont {J.}~\bibnamefont {Nocedal}}, \ and\ \bibinfo
  {author} {\bibfnamefont {C.}~\bibnamefont {Zhu}},\ }\href@noop {} {\bibfield
  {journal} {\bibinfo  {journal} {{SIAM} Journal on Scientific Computing}\
  }\textbf {\bibinfo {volume} {16}},\ \bibinfo {pages} {1190} (\bibinfo {year}
  {1995})}\BibitemShut {NoStop}%
\bibitem [{\citenamefont {Dennis}\ \emph {et~al.}(2013)\citenamefont {Dennis},
  \citenamefont {Hope},\ and\ \citenamefont {Johnsson}}]{XMDS}%
  \BibitemOpen
  \bibfield  {author} {\bibinfo {author} {\bibfnamefont {G.~R.}\ \bibnamefont
  {Dennis}}, \bibinfo {author} {\bibfnamefont {J.~J.}\ \bibnamefont {Hope}}, \
  and\ \bibinfo {author} {\bibfnamefont {M.~T.}\ \bibnamefont {Johnsson}},\
  }\href@noop {} {\bibfield  {journal} {\bibinfo  {journal} {Computer Physics
  Communications}\ }\textbf {\bibinfo {volume} {184}},\ \bibinfo {pages} {201}
  (\bibinfo {year} {2013})}\BibitemShut {NoStop}%
\end{thebibliography}%

\end{document}